\documentclass[12pt,preprint]{aastex}
\newcommand\kms{km s$^{-1}$}
\newcommand\msun{\ifmmode{M_{\odot}}\else $M_{\odot}$\fi}
\newcommand\rsun{\ifmmode{R_{\odot}}\else $R_{\odot}$\fi}
\newcommand\rgc{$R_{\rm GC}$}
\newcommand{\teff}{$T_{\rm eff}$} 

\begin{document}

\title{Elemental Abundance Ratios in Stars of the Outer Galactic Disk.\ III.\ 
Cepheids\footnote{This
paper makes use of observations obtained at the National Optical Astronomy
Observatory, which is operated by AURA, Inc., under contract from
the National Science Foundation. We also employ data products
from the Two Micron All Sky Survey, which is a joint project
of the University of Massachusetts and the Infrared Processing
and Analysis Center/California Institute of Technology, funded by the
National Aeronautics and Space Administration and the National
Science Foundation}}
\author{David Yong}
\affil{Department of Physics \& Astronomy, University of North
Carolina, Chapel Hill, NC 27599-3255; email: yong@physics.unc.edu}

\author{Bruce W.\ Carney}
\affil{Department of Physics \& Astronomy, University of North
Carolina, Chapel Hill, NC 27599-3255; email: bruce@physics.unc.edu}

\author{Maria Lu\'{i}sa Teixera de Almeida}
\affil{Department of Physics \& Astronomy, University of North
Carolina, Chapel Hill, NC 27599-3255; email: luisa@oal.ul.pt}

\author{Brian L.\ Pohl}
\affil{Department of Physics \& Astronomy, University of North
Carolina, Chapel Hill, NC 27599-3255; email: bpohl@physics.unc.edu}

\begin{abstract}

We present metallicities, [Fe/H], and elemental abundance ratios, [X/Fe], 
for a sample of 24 Cepheids in the outer Galactic disk based on high-resolution
echelle spectra. The sample have Galactocentric distances covering  
12 $\le$ \rgc~(kpc) $\le$ 17.2 making them the most distant Galactic Cepheids 
upon which detailed abundance analyses have been performed. We find sub-solar 
ratios of [Fe/H] and overabundances of [$\alpha$/Fe], [La/Fe], and [Eu/Fe]
in the program stars. All abundance ratios exhibit a dispersion that exceeds 
the measurement uncertainties. As seen in our previous studies of old open
clusters and field giants, enhanced ratios of [$\alpha$/Fe] and [Eu/Fe] 
reveal that recent star formation has taken place in the outer disk with Type 
II supernovae preferentially contributing ejecta to the ISM and with Type Ia 
supernovae playing only a minor role. The enhancements 
for La suggest that AGB stars have contributed to the chemical evolution of
the outer Galactic disk. 

Some of the young Cepheids are more 
metal-poor than the older open clusters and field stars 
at comparable Galactocentric distances.
This demonstrates that the outer disk is not the end result of the 
isolated evolution of an ensemble of gas and stars. We showed previously 
that the older open clusters
and field stars reached a basement metallicity at about 10-11 kpc. The younger
Cepheids reach the same metallicity but at larger Galactocentric distances, 
roughly 14 kpc. This suggests that the Galactic disk has been 
growing with time as predicted from numerical simulations.

The outer disk Cepheids appear to
exhibit a bimodal distribution for [Fe/H] and [$\alpha$/Fe]. Most of the
Cepheids continue the trends with Galactocentric distance 
exhibited by Andrievsky's larger Cepheid sample and we refer to these 
stars as the ``Galactic Cepheids''. A minority of the Cepheids show
considerably lower [Fe/H] and higher [$\alpha$/Fe] and we refer to these 
stars as the ``Merger Cepheids''. One signature of a merger event would 
be composition differences between the ``Galactic'' and ``Merger'' Cepheids. 
The Cepheids satisfy this requirement and we speculate
that the distinct compositions suggest that the ``Merger Cepheids'' may have
formed under the influence of significant merger or accretion events. 
The short lifetimes of the Cepheids reveal that the merger event may be
on-going with the Monoceros ring and Canis Major galaxy being possible  
merger candidates. 

\end{abstract}


\section{INTRODUCTION}

The Galactic disk accounts for the vast majority of stars in the Milky Way. 
Abundance analyses of
nearby field stars \citep{bdp93,bdp03}, open clusters \citep{friel95}, 
planetary nebulae \citep{henry04}, and H II regions \citep{shaver83} 
have provided insight into the mean metallicity and metallicity gradient 
of younger and older stars and stellar remnants. 
Radial abundance gradients, as measured in disk stars, provide crucial 
constraints for models of the formation and evolution of our Galaxy 
\citep{hou00,chiappini01}. However, to explore the origin and continuing 
evolution of the Galactic disk, we need much more information than mean 
metallicities alone can provide. We need detailed elemental abundance 
ratios, which contain vital information on the relative contributions of Type
II supernovae, Type Ia supernovae, and asymptotic giant branch (AGB) stars. 

In order to address this situation, we commenced an observing program to
measure metallicities and detailed chemical compositions for stars in the
outer Galactic disk. An analysis of the radial velocities 
and chemical abundance patterns of four old open clusters with 
Galactocentric distances between 12 and 23 kpc was presented in Paper I 
\citep{open}. In Paper II \citep{warp}, we continued our attack upon the outer
disk by initiating a successful selection process for identifying distant 
field giants then conducting an abundance analysis of three such stars. 
From these analyses, two principal findings may be noted: (1) at large 
Galactocentric distances, \rgc~$> 10$ kpc, 
the expected metallicity gradient
vanishes and the stars exhibit a constant value of [Fe/H] $\approx$ $-$0.5; 
(2) field and cluster stars in the outer Galactic disk show enhancements for 
the ``$\alpha$'' elements, [$\alpha$/Fe] $\approx$ 0.2. Our interpretation 
was that these abundance patterns reflected the episodic growth
of the disk via accretion or merger events. These events triggered rapid star 
formation with Type II supernovae preferentially contributing to the chemical 
enrichment. 

The open clusters are old, with ages between 2 - 6 Gyr. The ages of the field 
giants are unknown, but presumably these stars are quite old too. 
The absence of the radial abundance gradient inferred from our 
sample of distant field and clusters stars may therefore not reflect the 
current situation in the outer disk. 
The time variation of the Galactic radial abundance gradient offers a more 
comprehensive test of Galactic evolution models than a single epoch 
(or time integrated) abundance
gradient. The possibility of an on-going merger event in the outer disk 
(e.g. \citealt{newberg02,ibata03,yanny03,martin04}) 
reinforces the need to analyze a sample of 
young stars at large distances and to measure detailed abundance 
ratios [X/Fe] in addition to the metallicity, [Fe/H]. 

Cepheids are high-mass stars whose short lifetimes ensure that their 
atmospheres reflect the present-day composition of the ISM. Abundance
analyses of Cepheids 
appear feasible 
\citep{fry97,andrievsky02a,andrievsky02b,andrievsky02c,andrievsky04,luck03} 
and the Cepheid period-luminosity relation allows for accurate distance 
determinations. Due to their luminosity, high-resolution spectroscopic
observations of
Cepheids located at large distances can be conducted with modest-sized 
telescopes. In this paper, the third of this series, 
we present metallicities, [Fe/H], and elemental 
abundance ratios, [X/Fe], for a sample of two dozen Cepheids 
with Galactocentric distances 12 $\le$ \rgc~$\le$ 17.2 
kpc. These Cepheids allow us to study the Galactic radial abundance
gradient as a function of time. Detailed abundance ratios [X/Fe] offer
an insight into the events currently taking place in the outer disk and
such abundance ratios may reveal the signatures of recent merger events. 

\section{PROGRAM STARS AND OBSERVATIONS}

\subsection{Target selection}

We compiled a list of 421 Type I Cepheids using three
primary sources: the Cepheid database maintained by
Fernie at the University of 
Toronto\footnote{www.astro.utoronto.ca/DDO/research/cepheids/cepheids.html},
\citet{caldwell87}, and \citet{metzger98}. 
Periods were taken from the Fourth Edition of the General Catalog
of Variable Stars\footnote{www.mai.sai.su/groups/cluster/gcvs/gcvs}.
Optical photometry and reddening estimates were obtained from these
references, supplemented by reddening estimates of Fernie (1990).
We then calculated heliocentric and Galactocentric distances for 
all these stars using the visual band period-luminosity relation 
of \citet{madore91}. We identified a list of thirty 
Cepheids with Galactocentric distances \rgc~$>$ 12.5 kpc of which
10 had \rgc~$>$ 14 kpc. 
Spectra for all 10 Cepheids with \rgc~$>$ 14 kpc
and 14 of the 20 with 12.5 $<$ \rgc~$<$ 14.0 kpc were obtained during 
a series of observing runs (see Table \ref{tab:basic}).

\subsection{Observations and data reduction}

The targets were observed using the echelle spectrographs on the 
4-meter telescopes at the Kitt Peak National Observatory (KPNO) and the Cerro 
Tololo Inter-American Observatory (CTIO) during four different observing runs
between 1997 December and 1999 January. Red long red cameras were used with 
the 31.6 lines~mm$^{-1}$ echelle gratings. Second-order blue light was blocked 
using GG495 filters. The wavelength coverage was 5500-8000\AA~at CTIO and
4700-8000 \AA~at KPNO with the difference resulting from the choice of 
cross-dispersers, G181 at CTIO (316 lines~mm$^{-1}$) and G226 at KPNO 
(226 lines~mm$^{-1}$). A 1.0\arcsec\ slit (150 microns) provided a spectral
resolution of 28,000 (two pixels per resolution element) and a dispersion of
0.07 pixel at 5800\AA. Generally we were able to achieve signal-to-noise 
ratios (S/N) of about 75 per pixel (108 per resolution element). 

\begin{figure}
\epsscale{0.9}
\plotone{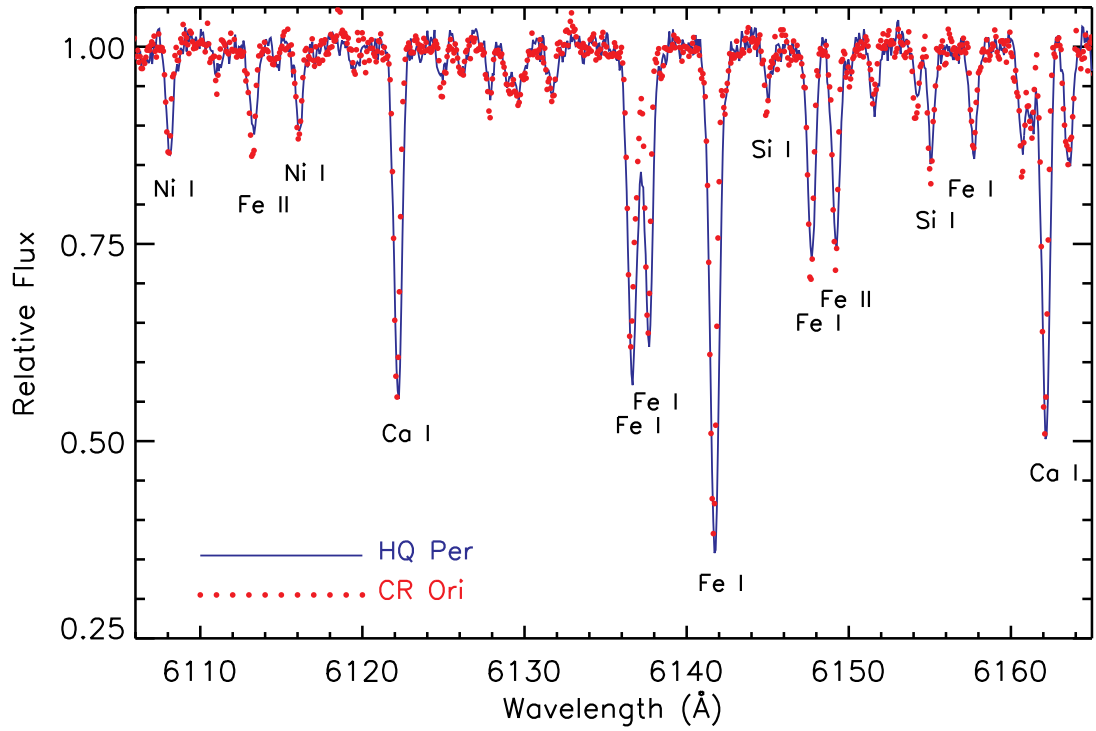}
\caption{Spectra of HQ Per (\rgc~= 13.3 kpc) and CR Ori (\rgc~= 13.2 kpc) 
between 6110 and 6160\AA.
The spectra are very similar suggesting that these Cepheids likely
have comparable stellar parameters and abundances. 
\label{fig:spectra}}
\end{figure}

\begin{figure}
\epsscale{0.8}
\plotone{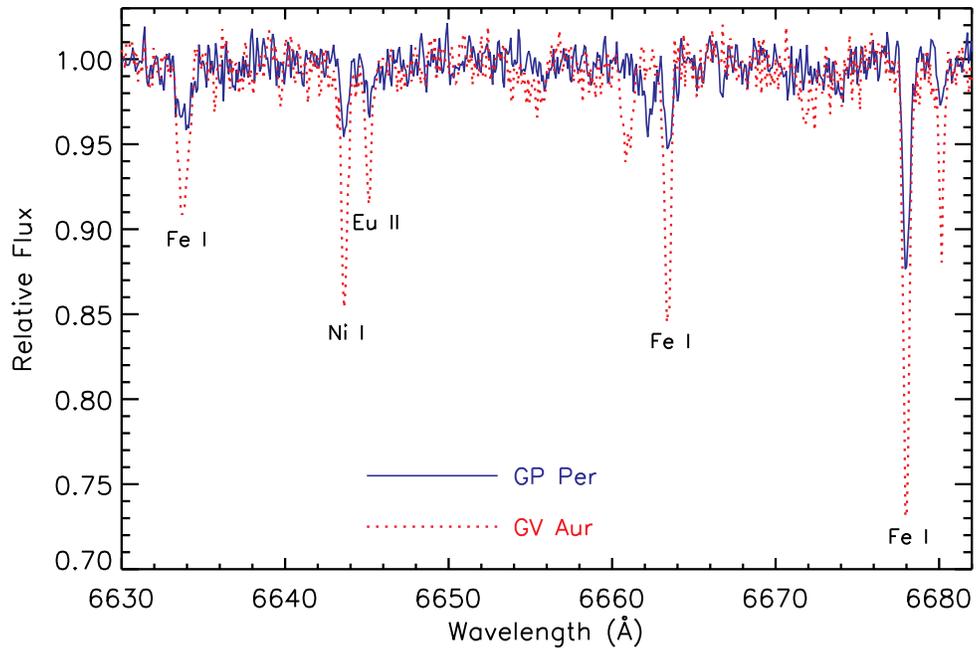}
\caption{Spectra of GP Per and GV Aur near the 6645\AA~Eu line.
The Cepheids have identical effective temperatures \teff~= 6400K
so the contrasting line strengths reflect real abundance differences. 
\label{fig:eu}}
\end{figure}

The observing routine included 20 quartz lamp exposures
to provide data for flat-fielding, and 15 zero-second
exposures to provide ``bias" frames. Th-Ar hollow
cathode lamp spectra were taken before and after each
stellar exposure. The data were reduced using the 
IRAF\footnote{IRAF (Image Reduction and Analysis
Facility) is distributed by the National Optical Astronomy
Observatory, which is operated by the Association of Universities
for Research in Astronomy, Inc., under contract with the National
Science Foundation.} packages IMRED, CCDRED, and ECHELLE to correct
for the bias level, trim the overscan region, extract individual
orders, fit the continuum, apply a wavelength solution using
the Th-Ar spectra. Master flat field frames
were produced each night, and normalized using APFLATTEN, following
which the data frames were divided by the master flat field frames
prior to extraction of individual orders using APALL. The wavelength
solution was determined from Th-Ar comparison spectra obtained 
after each program star. ECIDENTIFY and DISPCOR were used to identify 
the Th-Ar lines and
determine the dispersion solution for each order, and the CONTINUUM
task enabled us to interactively fit a high-order cubic spline
to produce the continuum-normalized, wavelength-calibrated
spectra. Stars with more than one observed spectrum were 
cross-correlated and then
combined into a single final spectrum using SCOMBINE. 
(The spectra were only combined if they were obtained sufficiently 
close in time to ensure that the phases were essentially identical.) 
In Figures \ref{fig:spectra} and \ref{fig:eu}, we show the final reduced 
spectra for two pairs of Cepheids. 
As we describe below, we believe that the two Cepheids in Figures 
\ref{fig:spectra} and \ref{fig:eu} have nearly identical effective 
temperatures ($\approx$ 5625K in Figure \ref{fig:spectra} and 
$\approx$ 6400K in Figure \ref{fig:eu}). Figure \ref{fig:eu} clearly shows 
that there is a significant spread in chemical abundances among Cepheids in 
the outer Galactic disk. 

\subsection{Overtone pulsators}

Most Cepheids pulsate in the fundamental mode, and the
period-luminosity relation applies only to such stars.
Nonetheless, we must be careful to try to verify
the pulsational mode of our program stars so that we
may reliably calculate their distances. 

First overtone pulsators are more common among the
shorter period stars. For stars with metallicities
similar to what are expected in the outer Galactic
disk, and which we obtain for our program stars,
an excellent reference is the very extensive work
by \citet{alcock95} on Cepheid variables in
the Large Magellanic Cloud. Their Figure~5 shows
clearly the offset in the $V$ vs.\ log~$P$ relation
that identifies the overtone pulsators. It is clear
that at the metallicity of the LMC, [Fe/H] $\approx$ $-0.3$,
the stars with periods shorter than about 2.5 days have
a reasonably high probability of pulsating in the
first overtone mode.

We have relied upon the on-line version of the General
Catalog of Variable 
Stars 
to help identify the first overtone pulsators in
our sample, which may generally be identified by the
more sinusoidal, more symmetric shape of the optical
light curves. NY~Cas is pulsating in the first overtone,
but we note that Table~1 reveals that three other stars,
EW~Aur, GP~Per, and XZ~CMa, have pulsational periods of
less than 3 days. We have corrected the observed pulsation
period of NY~Cas to its equivalent fundamental mode, 4.01 days,
using the relation provided by \citet{alcock95}. We have not adjusted 
the distance estimates for the other three stars, but caution the 
reader that these distances may be underestimated. 

\subsection{Distance estimations}

\begin{figure}
\epsscale{0.8}
\plotone{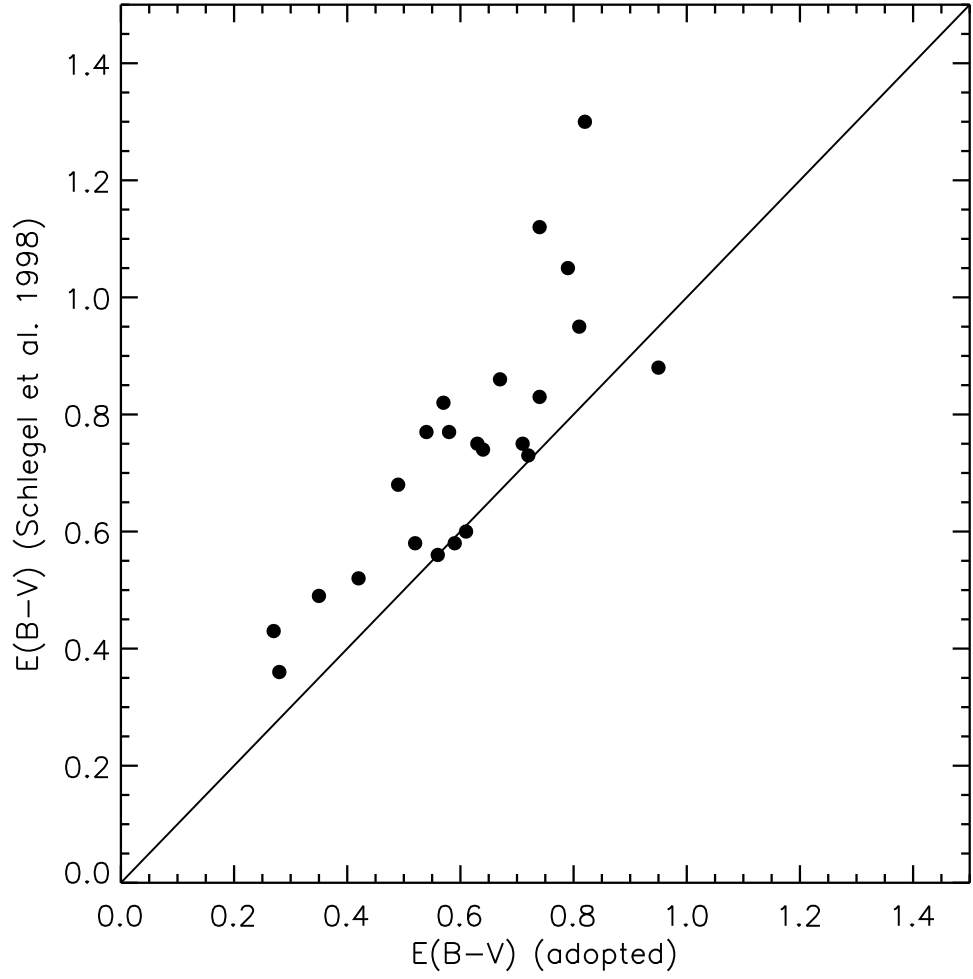}
\caption{Comparison of the reddening E(B-V) between our adopted
values and those from \citet{schlegel98}. \label{fig:redd}}
\end{figure}

Distances to Cepheids are readily obtained from the period-luminosity 
relation. Our program stars were selected to lie at large Galactocentric 
distances and are located in the Galactic plane. Therefore the
effects of interstellar reddening and extinction are important and must be
taken into account. At longer
wavelengths, extinction and reddening are 
less significant such that $A_K = 0.11A_V$. 
Therefore, we rely on infrared photometry from 2MASS to determine the distances 
to the program Cepheids. We employ the relation from \citet{madore91} 
\begin{equation}
\label{eq:mk}
M_{K({\rm CIT})} = -3.42 (\log P - 1.00) - 5.70
\end{equation}
after converting the magnitudes from the 2MASS system to the CIT system
via the transformations provided by \citet{carpenter01}. We also converted 
the single
epoch 2MASS $K$ magnitudes to mean $K$ magnitudes according to the prescription
given by \citet{soszynski05}. 
In Figure \ref{fig:redd} we compare our adopted 
reddenings with those derived from the 
\citet{schlegel98} dust maps. As expected, the
adopted reddenings are all equal to or less than the \citet{schlegel98}
values. Our program Cepheids are located at Galactocentric
distances 12.0 $\le$ \rgc(kpc) $\le$ 17.2. 
Seven of our Cepheids, CE Pup, EE Mon, ER Aur, 
FO Cas, HQ Car, IO Cas, and NY Cas, 
lie at Galactocentric distances beyond \rgc~= 15 kpc and are the
most distant Galactic Cepheids whose chemical compositions have been analyzed. 

We assert that the largest uncertainty in our 
distance estimation is due to the conversion from 
single epoch 2MASS K magnitudes to mean K magnitudes. \citet{soszynski05} 
estimate that the uncertainties in mean K magnitude for Galactic Cepheids 
are about 0.03 mag which corresponds to distance errors of 0.2 kpc. 
A 0.1 mag error in E($B-V$) also corresponds to a distance error of 0.2 kpc. 
Perhaps the most direct test of the uncertainties in our distances 
is to compare the values derived from $V$ and from $K$. The mean difference 
is distance (K) $-$ distance (V) = 0.23 kpc ($\sigma$ = 0.72). 
Therefore we take 0.7 kpc as a representative uncertainty 
in our Cepheid distances. 

\section{ELEMENTAL ABUNDANCES}

\subsection{Stellar parameters}

The effective temperature (\teff), surface gravity ($\log g$), and
microturbulent velocity ($\xi_t$) were determined using spectroscopic
criteria. This is the same method that was used in Paper I and Paper II
as well as by \citet{fry97} who undertook an analysis of nearby Cepheids that
calibrate the period-luminosity relation. 
Equivalent widths were measured for a set of 
Fe\,{\sc i} and Fe\,{\sc ii} lines using routines in IRAF. The $gf$-values
for the Fe\,{\sc i} lines were taken from the laboratory measurements 
performed by the Oxford group 
(e.g., \citealt{blackwell79feb,blackwell79fea,blackwell80fea,
blackwell86fea,blackwell95fea} and references therein). For Fe\,{\sc ii} 
lines, we used the $gf$-values from \citet{biemont91}. The full list
of Fe\,{\sc i} and Fe\,{\sc ii} lines are presented in Table \ref{tab:line}. 
Local thermodynamic equilibrium (LTE) model atmospheres were computed using 
the ATLAS9 program \citep{atlas9} and we used the 
LTE stellar line analysis program {\sc Moog} \citep{moog}. We set
\teff~by forcing the abundances from Fe\,{\sc i} lines to be independent
of the lower excitation potential. We adjusted $\log g$ until the abundances
from Fe\,{\sc i} and Fe\,{\sc ii} agreed. Finally, the microturbulence was
determined by insisting that the abundances from Fe\,{\sc i} lines show
no trend versus EW. This process required iteration until a self-consistent
set of parameters was obtained. In Papers I and II we employed an identical 
procedure for determining the stellar parameters of the older cluster and 
field red giants. In Section \ref{sec:valid} below 
we consider more carefully whether our approach is appropriate. 
The stellar parameters presented in Table
\ref{tab:param} are those adopted for the subsequent abundance analysis. 

\subsection{Elemental abundance analysis}

Next we measured EWs for lines of the $\alpha$ 
elements Mg, Si, Ca and Ti again using routines
in IRAF. Abundances were determined via {\sc Moog} based on the measured EW,
model atmosphere, and atomic data. 
The $gf$-values for Ca and Ti were taken from the Oxford group 
\citep{smith81ca,blackwell82tia,blackwell83tia,blackwell86tia}. For Si,
an inverted solar analysis assuming log~$\epsilon$(Si)~=~7.55 was used to
determine the $gf$-values. For Mg, we relied upon the $gf$-values employed by 
\citet{ramirez02}. For lines of the $s$-process element La, 
we took the $gf$-values from \citet{la} and included the effects of 
hyperfine splitting. 
For the $r$-process element Eu, the $gf$-values were taken from \citet{eu} and 
take into account hyperfine and isotopic splitting (we assumed a solar 
isotopic mix). To measure the Eu abundance, we performed spectrum synthesis. 
The full list of atomic lines is presented in Table \ref{tab:line} and
the adopted solar abundances were presented in Paper I. 
Recall that in Paper I our line lists and analysis procedures 
successfully reproduced the abundance distributions in the Sun, Arcturus,
and three stars close to the tip of the red giant branch in the old
open cluster 
M67. We present the abundance ratios for the program Cepheids in Table 
\ref{tab:abund}. Typical internal errors for our spectroscopic model
parameters are \teff~$\pm$ 150K, $\log g$~$\pm$ 0.3 dex, and $\xi_t$ $\pm$
0.3 \kms. In Table \ref{tab:err} we present the abundance dependences
upon the model parameters. 

\subsection{Comparison with Fry \& Carney}

\begin{figure}
\epsscale{0.8}
\plotone{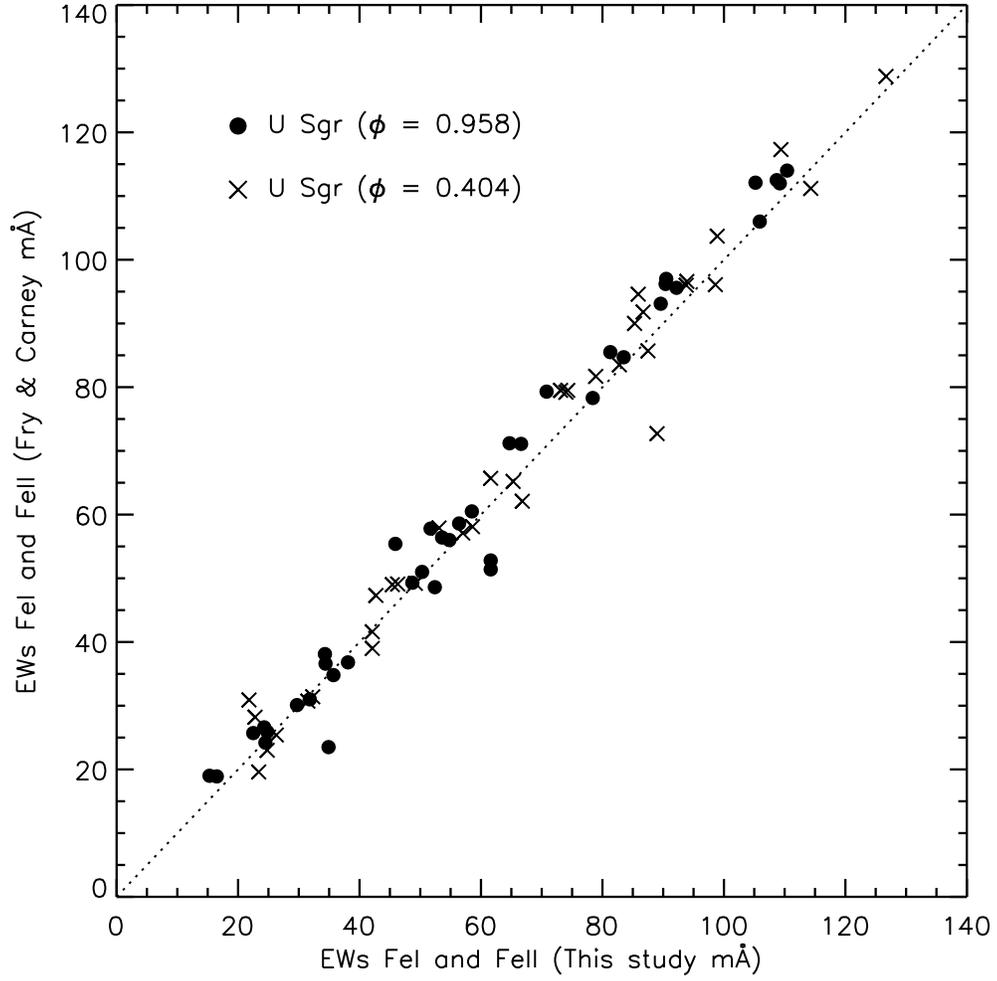}
\caption{Comparison of the EWs of Fe\,{\sc i} and Fe\,{\sc ii} lines 
between this study and \citet{fry97} for U Sgr at two different
phases. \label{fig:ew_comp}}
\end{figure}

\begin{figure}
\epsscale{0.8}
\plotone{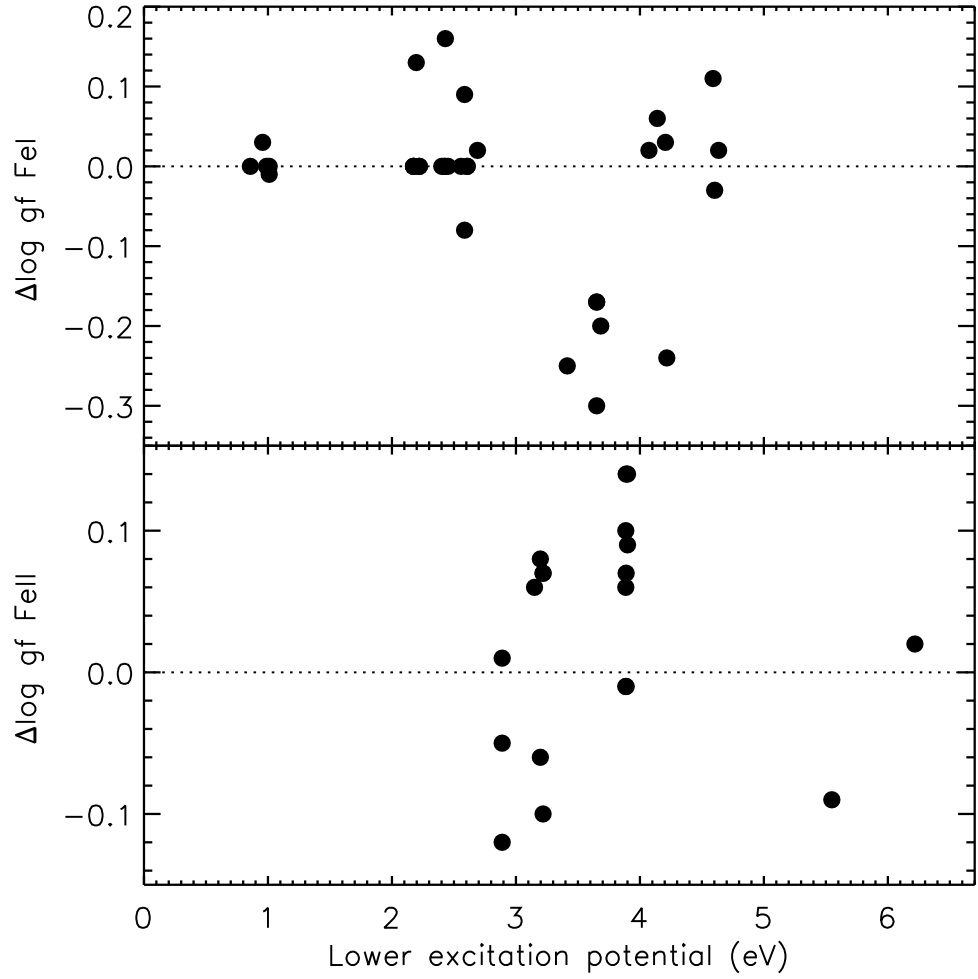}
\caption{Comparison of the $gf$-values of Fe\,{\sc i} (upper) and Fe\,{\sc ii} 
(lower) lines between this study and \citet{fry97}. \label{fig:gf_comp}}
\end{figure}

While the program stars are outer disk Cepheids, our overall goal is to 
compare the young stars in the inner and outer Galactic disk. 
In order to compare the compositions of the outer 
disk Cepheids with their solar neighborhood 
counterparts, we re-analyzed a subset of the \citet{fry97} sample that 
included Cepheids with 6 $\le$ \rgc~$\le$ 10 kpc. The 
principal differences between these two analyses are the adopted $gf$-values 
for Fe as well as the selection of Fe lines. The first simple test 
performed concerns the measurement of EWs in identical
spectra. For U Sgr at two different phases, we 
measured EWs for the lines analyzed by \citeauthor{fry97}. In Figure 
\ref{fig:ew_comp}, we compare the measured EWs. For the 
75 common lines, the mean difference for {\sc this study $-$ fry \& carney} 
is $-$1.6 $\pm$ 0.5 ($\sigma$ = 4.6 m\AA). Such a 
comparison demonstrates that the measured 
EWs cannot be responsible for any systematic differences that may arise 
between the two studies. Next we consider the $gf$-values used by the two 
studies. In Figure \ref{fig:gf_comp}, we compare the adopted $gf$-values for 
the lines common to both studies. The scatter for Fe\,{\sc i} lines is
considerable, especially for lines with lower excitation potentials between
2 and 5 eV with
the maximum discrepancy reaching 0.3 dex. Not all lines
are measurable in every Cepheid. Therefore, depending on the set of measured 
lines, we anticipate possible differences in \teff~between the two studies 
resulting from the adopted Fe\,{\sc i} $gf$-values. We may also expect some
differences in microturbulent velocities. 
The $gf$-values for Fe\,{\sc ii} lines
are in good agreement with the maximum difference being about 0.15 dex. 
Depending on the selection of measured Fe\,{\sc ii} lines, there may be 
small differences in the derived surface gravity. 

\begin{figure}
\epsscale{0.8}
\plotone{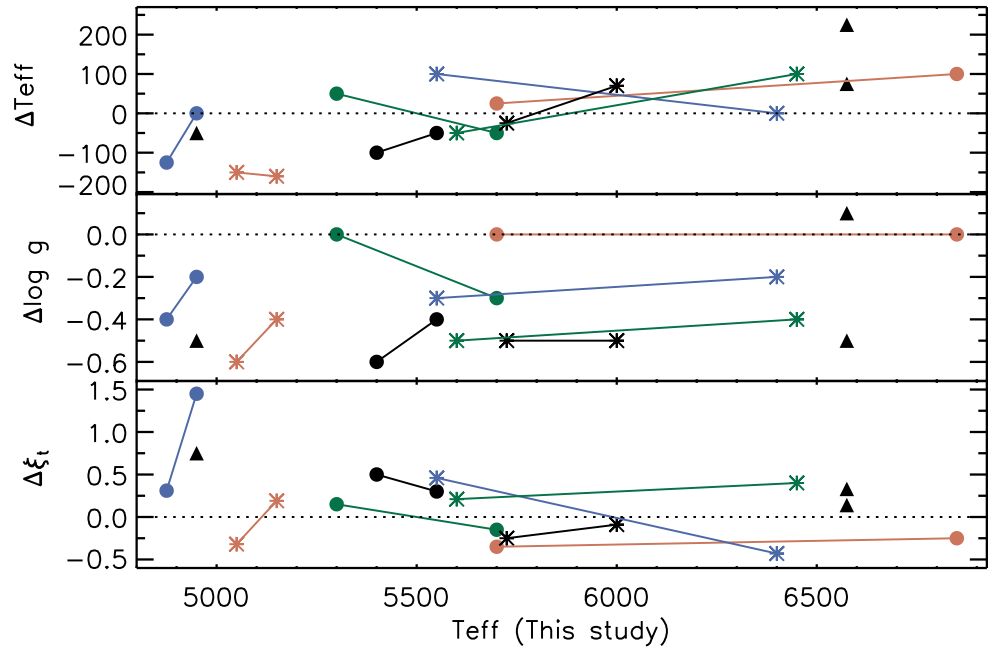}
\caption{Stellar parameter differences for \teff, $\log g$, and $\xi_t$ 
for {\sc this study $-$ fry \& carney} versus \teff~(this study). Solid lines 
connect the same Cepheids observed at different phases. 
\label{fig:fry.param}}
\end{figure}

For the comparison sample of local Cepheids, we 
utilize the identical spectra obtained and analyzed by 
\citeauthor{fry97}. Recall that \citeauthor{fry97} applied the same 
spectroscopic criteria that we have imposed when setting their stellar 
parameters. In Table \ref{tab:fry} we show our stellar parameters and abundance 
ratios for a sample of solar neighborhood Cepheids. In total there are 
11 Cepheids observed at 19 phases in common between the two studies. 
In Figure \ref{fig:fry.param} we plot the differences in stellar 
parameters $\Delta$\teff, $\Delta\log g$, and $\Delta\xi_t$ versus
\teff~(this study). 
The mean offsets are $\Delta$\teff~ = $-$1 $\pm$ 23 ($\sigma$ = 100~K),
$\Delta\log g$ = $-$0.33 $\pm$ 0.05 ($\sigma$ = 0.22 cgs), 
and $\Delta\xi_t$ = 0.18 $\pm$ 0.10 ($\sigma$ = 0.45 \kms). The quoted
differences are for {\sc this study $-$ fry \& carney}.
An interesting feature is that there appears to be a trend
between $\Delta$\teff~and \teff~(this study). This may be related to the 
difference in $gf$-values between the two studies. 
Another possibility is that lines 
with systematically or randomly different $gf$ values 
at lower and higher excitation potentials are being used. 
Our gravities are systematically lower than those obtained
by \citeauthor{fry97}. 
In this case, there doesn't appear to be a trend with \teff. The difference
in $\log g$ may also be a result of line selection and the corresponding 
$gf$-values. For the microturbulence, there is one obvious outlier. If we
remove this outlier, there does not appear to be a trend between 
$\Delta \xi_t$ and \teff. For all stellar parameters, we emphasize that the 
mean offsets and dispersions are comparable to the internal uncertainties
of this study. 

\begin{figure}
\epsscale{0.8}
\plotone{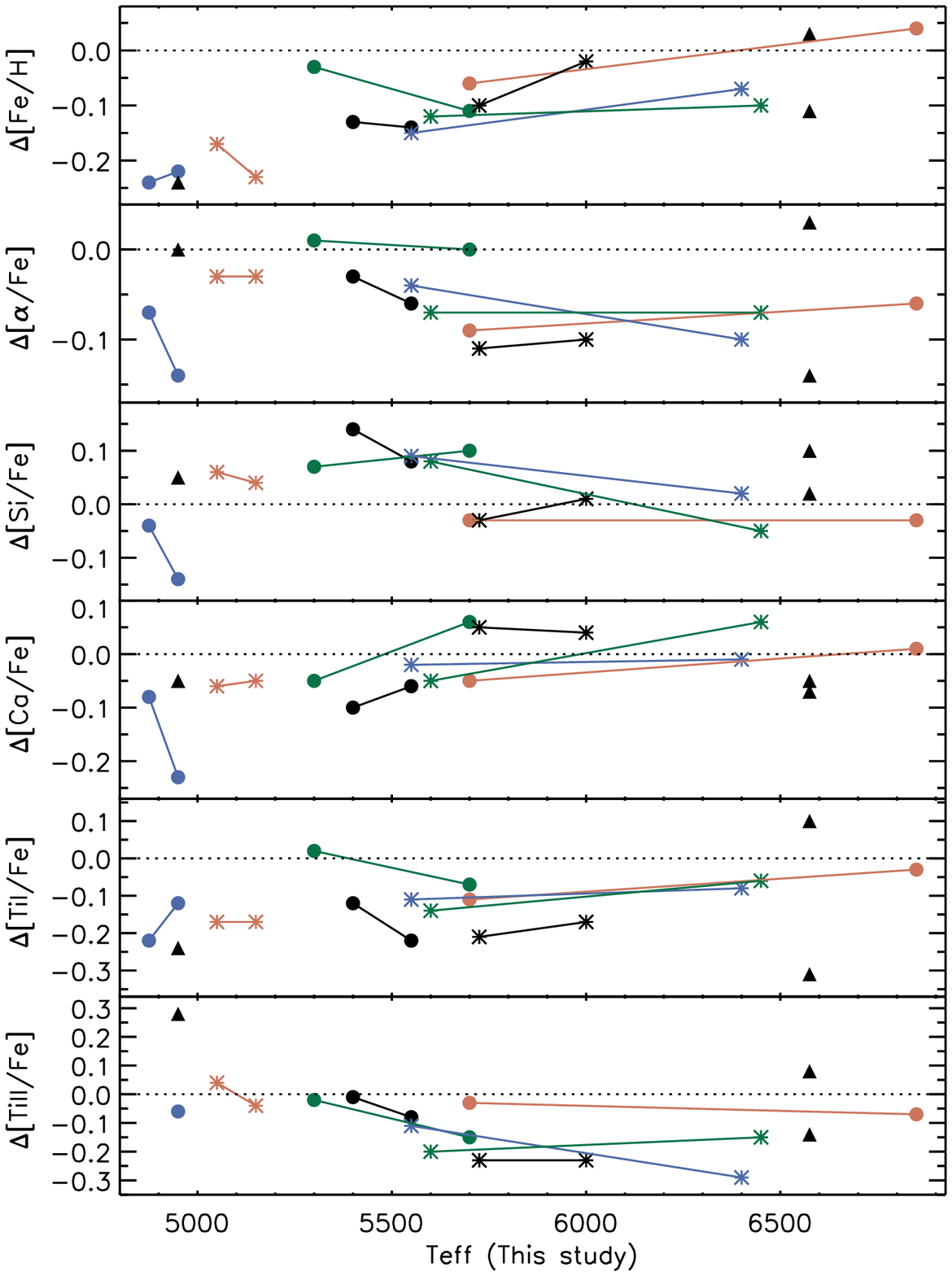}
\caption{Abundance differences for [Fe/H] and [X/Fe] for {\sc this 
study $-$ fry \& carney} versus \teff~(this study). Solid lines 
connect the same Cepheids observed at different phases. 
\label{fig:fry.abund}}
\end{figure}

For the purposes of this study, the crucial offsets we seek to measure are
for the abundance ratios [Fe/H] and [X/Fe]. 
In Figure \ref{fig:fry.abund}, we plot the abundance
differences $\Delta$[Fe/H] and $\Delta$[X/Fe] for 
{\sc this study $-$ fry \& carney} versus \teff~(this study). The mean offsets
are 
$\Delta$[Fe/H] = $-$0.11 $\pm$ 0.02 ($\sigma$ = 0.08), 
$\Delta$[Si/Fe] = 0.03 $\pm$ 0.02 ($\sigma$ = 0.07), 
$\Delta$[Ca/Fe] = $-$0.04 $\pm$ 0.02 ($\sigma$ = 0.07), 
$\Delta$[Ti\,{\sc i}/Fe] = $-$0.13 $\pm$ 0.02 ($\sigma$ = 0.10), 
$\Delta$[Ti\,{\sc ii}/Fe] = $-$0.08 $\pm$ 0.03 ($\sigma$ = 0.13), and
$\Delta$[$\alpha$/Fe] = $-$0.06 $\pm$ 0.01 ($\sigma$ = 0.05).

The first point we draw attention to is that we confirm the 
\citeauthor{fry97} conclusion that self-consistent results can be obtained 
from a traditional spectroscopic analysis. For 8 Cepheids, we analyzed
the spectra obtained at two different phases and found that [Fe/H] and 
[X/Fe] are essentially identical (see columns 6-14 of Table \ref{tab:fry}). 
In some cases, the constancy of the measured compositions prevails despite
a change of 1000~K in a star's effective temperature. 

An interesting feature of the abundance comparison is that there appears 
to be a trend between $\Delta$[Fe/H] and \teff~(this study). 
As \teff~decreases, the magnitude of the offset 
between the iron abundances increases with the maximum discrepancy reaching 
$-$0.25 dex at \teff~= 5000~K. This is unsurprising given that 
$\Delta$\teff~showed a trend with \teff~(this study) and in 
Table \ref{tab:err}, the Fe abundance based on Fe\,{\sc i} lines is 
rather sensitive to \teff. Again, we suggest that this trend may be
due to line selection and the corresponding $gf$-values. 
While a more detailed investigation is beyond the scope of the current paper,
clearly the trend between $\Delta$\teff~and \teff~as well as the
trend between $\Delta$[Fe/H] versus \teff~warrant further attention. 
We intend to re-analyze the entire \citeauthor{fry97} sample. Though we have 
confirmed that self-consistency in the measured compositions can be obtained 
at two different phases using traditional spectroscopic techniques, a 
re-analysis over the entire pulsation cycle is of great interest. 

Despite the trend between $\Delta$[Fe/H] and \teff~(this study),
for all other abundance ratios [X/Fe], we find no obvious trends between
$\Delta$[X/Fe] versus \teff. Inspection of Table \ref{tab:err}, reveals that
the mean offsets for [X/Fe] between these two analyses are comparable to the 
uncertainties in the model parameters. 

\subsection{Comparison with Andrievsky}
\label{sec:luckcomp}

The Cepheids listed in Table \ref{tab:fry} were also studied in the
series of papers by Andrievsky and collaborators 
\citep{andrievsky02a,andrievsky02b,andrievsky02c,luck03,andrievsky04}.
In addition to these nearby Cepheids, four outer disk Cepheids 
CU Mon, EE Mon, HW Pup, and WW Mon were also 
analyzed by Andrievsky. In total there are 15 Cepheids observed at 23 phases
in common between these studies and ours. 

Andrievsky's analysis differs from this study and from \citet{fry97} in
several ways. First, their $gf$-values were obtained using an inverted
solar analysis. Second, the \teff~scale was set using line depth ratios. 
Third, microturbulent velocities were obtained from Fe\,{\sc ii} lines only,
not Fe\,{\sc i} lines. 
Finally, while the gravity was set via ionization equilibrium, the adopted  
Fe\,{\sc i} abundance was not the mean abundance based on all Fe\,{\sc i} lines. 
Instead, the Fe\,{\sc i} abundance was inferred by plotting the abundance
versus EW for all Fe\,{\sc i} lines then extrapolating to an EW of 0~m\AA.

\begin{figure}
\epsscale{0.8}
\plotone{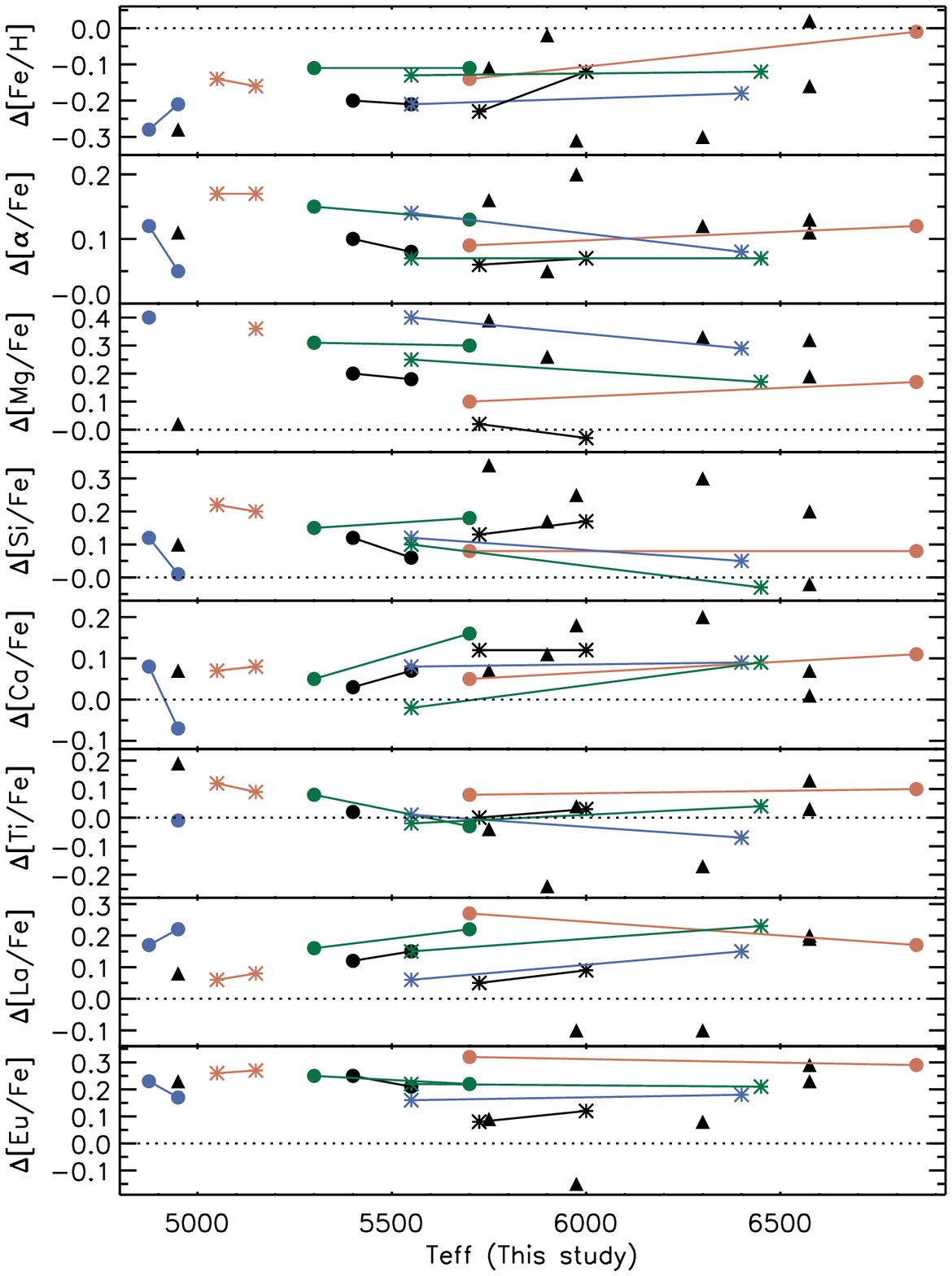}
\caption{Abundance differences for [Fe/H] and [X/Fe] for {\sc this 
study $-$ andrievsky} versus \teff~(this study). Solid lines 
connect the same Cepheids observed at different phases. 
\label{fig:luck.abund}}
\end{figure}

While there are 15 Cepheids in common, none were observed at identical phases.
Therefore we cannot directly compare the stellar parameters between the two
studies. However, we can compare
the derived abundance ratios [Fe/H] and [X/Fe]. In Figure \ref{fig:luck.abund} 
we plot the abundance differences $\Delta$[Fe/H] and $\Delta$[X/Fe] for
{\sc this study} $-$ {\sc andrievsky} 
versus \teff~(this study). The mean offsets are
$\Delta$[Fe/H] = $-$0.16 $\pm$ 0.02 ($\sigma$ = 0.09),
$\Delta$[Mg/Fe] = 0.23 $\pm$ 0.03 ($\sigma$ = 0.13),
$\Delta$[Si/Fe] = 0.13 $\pm$ 0.02 ($\sigma$ = 0.09),
$\Delta$[Ca/Fe] = 0.08 $\pm$ 0.01 ($\sigma$ = 0.06),
$\Delta$[Ti/Fe] = 0.02 $\pm$ 0.02 ($\sigma$ = 0.10),
$\Delta$[$\alpha$/Fe] = 0.11 $\pm$ 0.01 ($\sigma$ = 0.04), 
$\Delta$[La/Fe] = 0.12 $\pm$ 0.02 ($\sigma$ = 0.10), and
$\Delta$[Eu/Fe] = 0.19 $\pm$ 0.02 ($\sigma$ = 0.10). 

We again find a trend between $\Delta$[Fe/H] and \teff~as seen in the
comparison with \citeauthor{fry97}. Specifically, the magnitude of the
offset between iron abundances increases as \teff~decreases with
the maximum discrepancy reaching about $-$0.3 dex at \teff=5000~K. 
We believe this trend is due to the adoption by Andrievsky of the
\citet{fry97} effective temperatures to calibrate the line depth ratios
versus \teff~relation. 

There are no obvious trends between $\Delta$[X/Fe] and \teff,
despite the trend between $\Delta$[Fe/H] and \teff. This was also seen
in the comparison with \citeauthor{fry97} suggesting that abundance
ratios may be measured reliably. Again, we note that in many cases, the
mean offsets for [X/Fe] are comparable to the values arising from the
uncertainties in the stellar parameters (see Table \ref{tab:err}). 
In the subsequent sections, we apply these offsets to the 
Andrievsky Cepheid abundances when investigating the radial abundance
gradient. 

\subsection{Validity of a traditional spectroscopic approach}
\label{sec:valid}

Andrievsky chose a non-standard approach as described above 
(see \citealt{kovtyukh99} and \citealt{kovtyukh00} for a more detailed 
discussion). The fundamental reason for adopting a non-standard analysis 
was to avoid non-LTE effects. Fe\,{\sc i} lines are more susceptible than
Fe\,{\sc ii} lines to non-LTE effects due to the ionization balance in
F and G stars. That is, most of the Fe is in the form of Fe\,{\sc ii}
with Fe\,{\sc i} being the minor species at 
these temperatures. Overionization of Fe atoms
(relative to LTE) may result from the penetration by ultraviolet photons 
into the line-forming regions and such an effect would be lessened 
for the dominant species but may be serious for the minor component. 
In F and G stars, the effect of overionization
would be an underabundance of Fe\,{\sc i} in LTE analyses. As the dominant
species, Fe\,{\sc ii} would only be slightly affected and may show a small
overabundance. Were this a large
effect, it would compromise the \teff~scale based on excitation equilibrium
of Fe\,{\sc i}, the surface gravities based on ionization equilibrium of
Fe, and the microturbulent velocities set from Fe\,{\sc i} lines. 

Andrievsky's concerns regarding the reliability of the abundance analysis 
of Cepheids led to a non-traditional approach for the determination of the
stellar parameters. While this method has merit, we do have some 
concerns. We have noted that the temperatures 
used in Andrievsky's initial calibration 
of the line depth ratios were taken from the \citeauthor{fry97} \teff~scale. 
Recall that the \citeauthor{fry97} temperatures were 
derived using excitation equilibrium of Fe\,{\sc i} lines. Therefore, the 
very problem Andrievsky wished to avoid (i.e., possible non-LTE effects 
on Fe\,{\sc i} lines) was the underlying assumption that 
provided the effective temperatures for their initial calibration. 
Additionally, a given line depth ratio becomes increasingly uncertain 
as the S/N decreases. For the most distant Cepheids, Andrievsky et al.\
were working with S/N ratios as low as 40 from which 
accurate \teff~must be very difficult to measure from line depth
ratios.

If non-LTE effects of Fe\,{\sc i} lines are significant, then using 
Fe\,{\sc ii} lines to derive the microturbulence would appear to be a 
better method. However, for some program Cepheids, the small number of 
Fe\,{\sc ii} lines (e.g., N $<$ 10) may lead to less accurate values of 
$\xi_t$. 

\citet{luck85} recognized that the spectroscopically-determined gravity may 
systematically differ from the physical gravity in intermediate-mass
supergiants. A deficiency in the measured 
O abundances in Cepheids was noted and \citet{luck85}
showed that the O abundance was correlated with the difference between
physical and spectroscopic gravities. This correlation was interpreted
as a possible consequence of systematic errors in the analysis. 
An alternative explanation offered was that the O underabundance was 
already present in the interstellar gas out 
of which the stars formed. Andrievsky et al.\ 
found that the spectroscopic gravity can be made to match the 
physical gravity if the Fe\,{\sc i} abundance is estimated by extrapolating 
to an EW of 0~m\AA. In combination with their \teff~and microturbulent 
velocities, Andrievsky's gravity scale removed the O underabundance. 

\begin{figure}
\epsscale{0.8}
\plotone{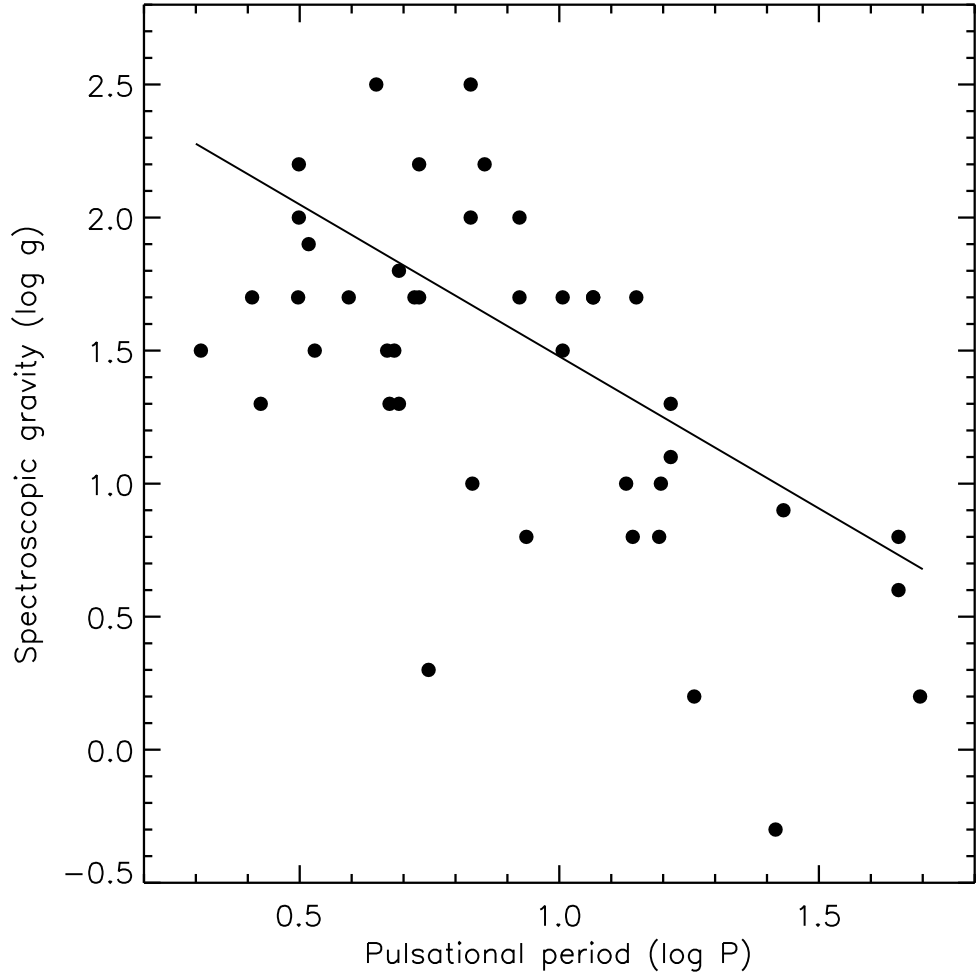}
\caption{Spectroscopic gravities versus pulsational periods for the outer disk
and comparison solar neighborhood Cepheids. The solid line
is the period-gravity relation for radially pulsating variable stars
defined by \citet{fernie.grav95}. 
\label{fig:period}}
\end{figure}

In Figure \ref{fig:period} we plot the spectroscopically determined gravities
versus pulsational period and overplot the period-gravity relation for variable
stars defined by \citet{fernie.grav95}. For the longer period Cepheids, the
spectroscopic gravities tend to be lower than the physical gravities. In
general,
the spectroscopic gravities tend to scatter about the period-gravity 
relation. For our programs Cepheids and the subset of the \citet{fry97} sample
that we re-analyzed, we find a mean difference $\Delta \log\ g$ = 
$\log g_{\rm spec} - \log g_{\rm period}$ = $-$0.19 $\pm$ 0.07 ($\sigma$ =
0.48). 
The scatter is considerably larger than our estimated internal error of about
0.3 dex, and the similar amount of scatter in the comparison with the results
from \citet{fry97}. We attribute the 0.5 dex scatter to three causes, in addition
to our internal uncertainties. First, of course, the actual $\log\ g$ vs.\
$\log\ P$ relation must have scatter. Unfortunately, the Cepheid portion of
the relation was defined using models rather than observations, so this
source of error cannot be quantified. Second, the distribution does not appear
to follow a simple Gaussian distribution. Three stars (IO~Cas, YZ~Aur, and
OT~Per) depart significantly from the relation. Excluding these three stars
lowers
the scatter in the predicted vs.\ derived $\log g$ values to only 0.40~dex,
which is probably consistent with the convolution of our uncertainties and
those in the relation of \citet{fernie.grav95}. Third, there is the question
of non-LTE effects noted above. Qualitatively, we would expect the longer
period
stars, with the lowest gravities, to be affected most. Indeed, two of the
three stars with the largest deviations have long periods, YZ~Aur (18.2 days)
and OT~Per (26.1 days). Other stars with comparably long periods, however,
appear to provide an excellent match to the relation. Because of the importance
of the non-LTE issue, we explore this further. 

\begin{figure}
\epsscale{0.8}
\plotone{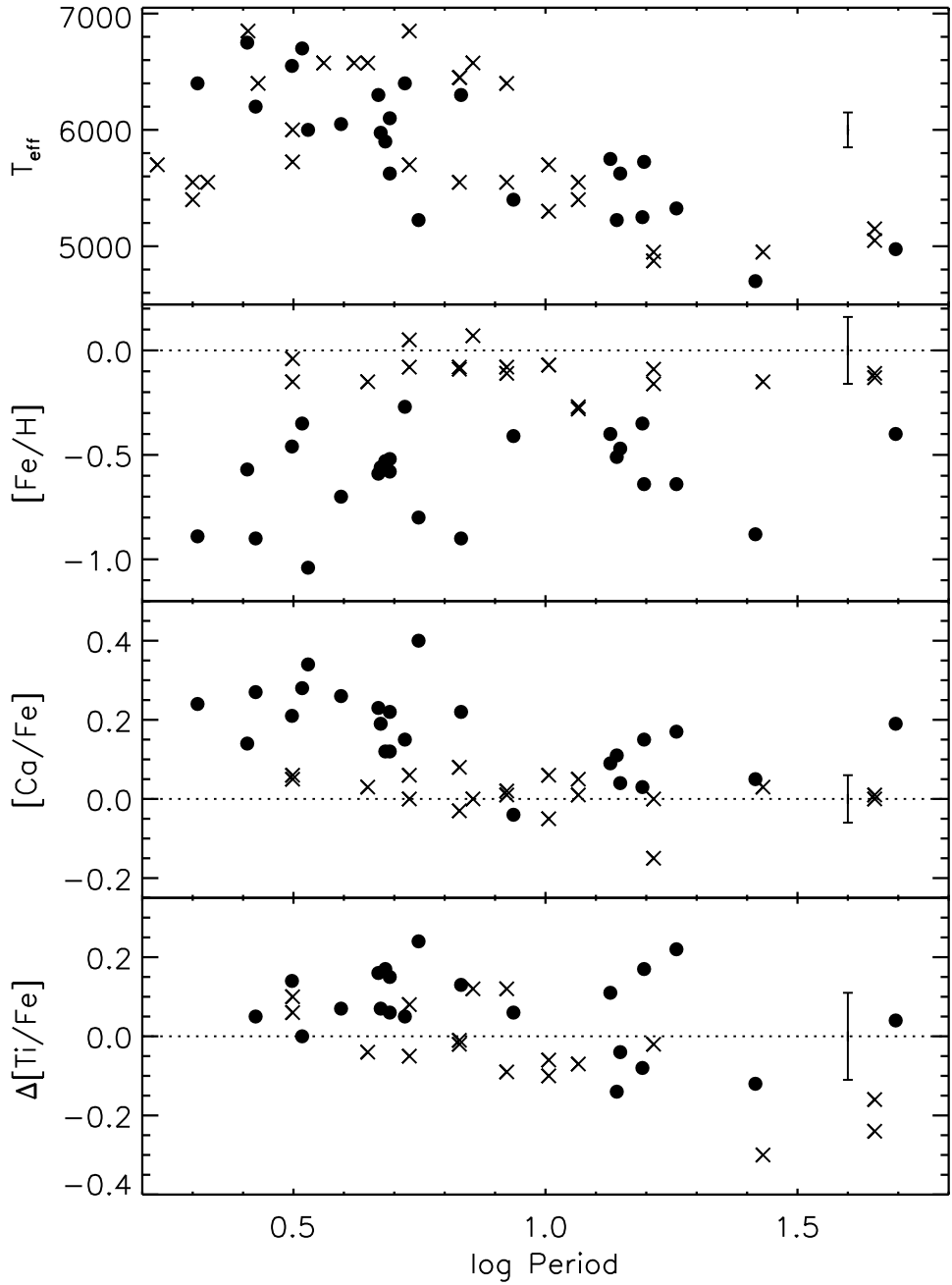}
\caption{\teff~and abundance ratios versus period. 
The closed
circles are our program Cepheids while the crosses represent the subset of
\citet{fry97} Cepheids re-analyzed in this study. A representative error 
bar is shown.
\label{fig:grav2}}
\end{figure}

We begin by considering the behavior of four of our key derived quantities
vs.\ the logarithm of the pulsation period
(see Figure \ref{fig:grav2}). We distinguish between stars from the
outer disk in this program, shown as filled circles, from those of the
local disk, from \citet{fry97} and re-analyzed by us (see Table~\ref{tab:fry}),
shown as crosses. The top panel shows that the temperatures derived
for the two samples share the same behavior as a function of period,
and consistent with the longer period, lower gravity stars in the
instability strip also being cooler. The second panel shows that the
local cepheids show very similar [Fe/H] values, independent of period
(and therefore gravity), while the outer disk cepheids are more metal-poor.
This foreshadows the following discussion, but is certainly expected in
the Galaxy due to an anticipated gradient in mean metallicity as a
function of Galactocentric distance. 

The third panel of Figure~\ref{fig:grav2} concentrates on possible
trends in element-to-iron ratios. We select calcium for the comparison because
the [Ca/Fe] values are well determined, and because its ionization potential
is the lowest of the four elements that define our measurable ``$\alpha$"
elements
(Mg, 7.65 eV; Ca, 6.11 eV; Si, 8.15 eV; and Ti, 6.82 eV). If non-LTE effects
are important, they should be more pronounced for elements with lower
ionization
potentials, and hence we might discover a trend in [Ca/Fe], since the
ionization
potential for iron is much higher, 7.90 eV. The Figure shows no such trend,
although it again foreshadows our results: the outer disk Cepheids show
enhanced
[Ca/Fe] values compared to local Cepheids. 

The bottom panel of Figure~\ref{fig:grav2} carries the non-LTE test further,
where
we compare the differences of [Ti/Fe] 
derived from lines of Ti\,{\sc i} and Ti\,{\sc ii}.
For the program Cepheids, we find a mean difference [Ti\,{\sc i}/Fe] $-$
[Ti\,{\sc ii}/Fe] = 0.07 $\pm$ 0.02 ($\sigma$ = 0.11). This difference
lies within the measurement uncertainties suggesting that the surface
gravities derived from ionization equilibrium of Fe are satisfactory.
There is a hint of a trend, however, in that the Cepheids with the
longest periods do appear to show a deficiency of [Ti\,{\sc i}/Fe]
relative to [Ti\,{\sc ii}/Fe]. This suggests that there may be a modest
degree of unaccounted-for over-ionization of Ti, presumably due to non-LTE,
but the effect is minor and appears to affect only the local Cepheids.

\begin{figure}
\epsscale{0.8}
\plotone{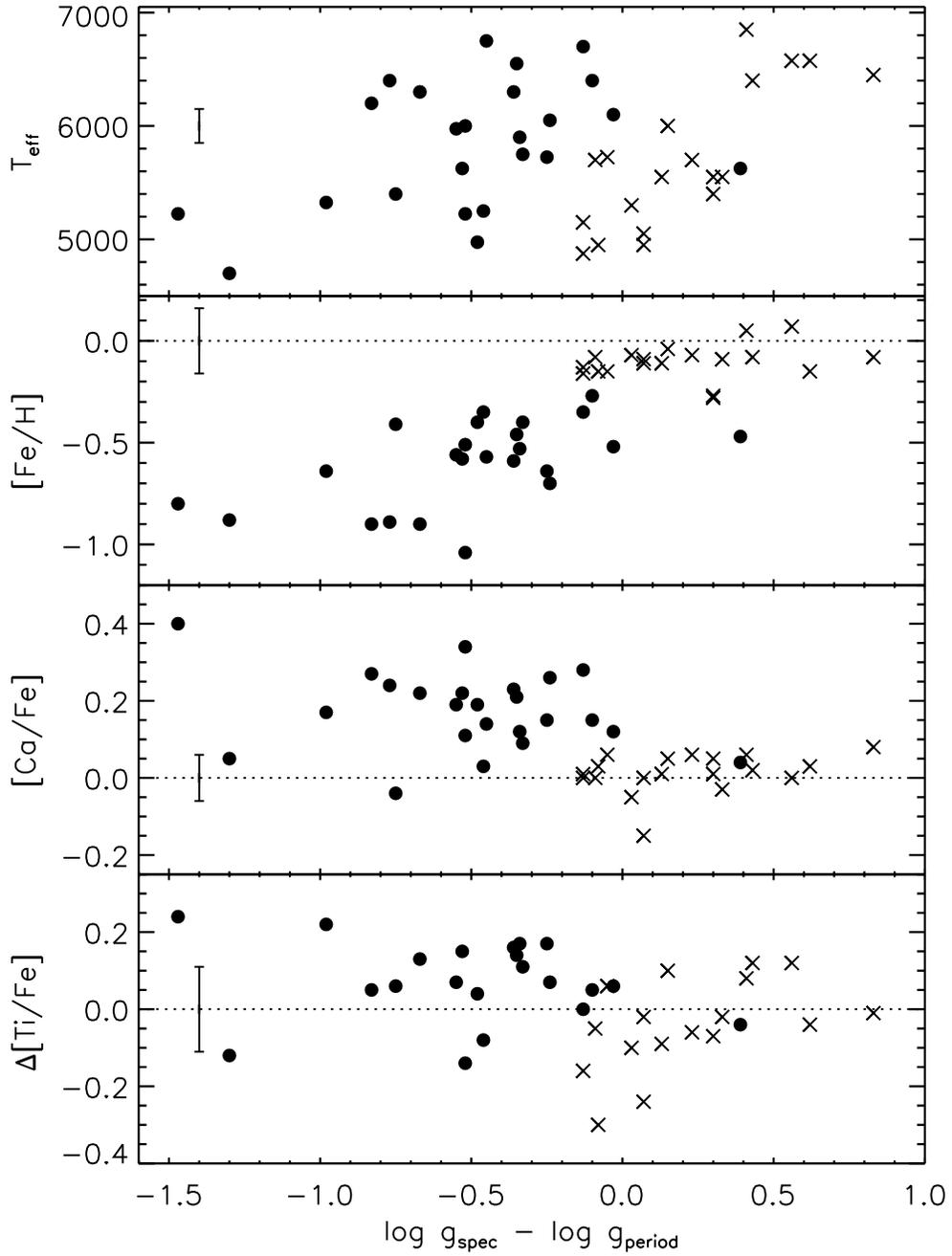}
\caption{\teff~and abundance ratios versus the difference in surface
gravity (spectroscopic $-$ period [\citealt{fernie.grav95}]). 
The symbols are the same
as in Figure \ref{fig:grav2}.
\label{fig:grav}}
\end{figure}

In Figure \ref{fig:grav}, we plot the same parameters against the
difference between spectroscopic and predicted gravities, ($\Delta \log g$). 
The three most significant outliers, with $\Delta \log\ g \leq\ -1$, are
IO~Cas ($-1.5$ dex), OT~Per ($-1.3$ dex), and YZ~Aur ($-1.0$ dex). Were IO~Cas
pulsating in the first overtone mode, the predicted fundamental period would
be longer, 8.06 days, but this would reduce the discrepancy only to $-1.3$ dex.
As noted above, the other two stars have relatively long periods, and the
extreme difference between the predicted and derived gravities may be due
to over-ionization, but, again, not all long period Cepheids show the
same behavior.

The trends in the four panels of Figure~\ref{fig:grav} reflect those
noted already in Figure~\ref{fig:grav2}. The [Fe/H], [Ca/Fe], and
$\Delta$[Ti/Fe] values of the outliers match those of the stars with smaller
differences (when allowances are made for comparing only outer disk Cepheids
amongst themselves, for reasons discussed above). The one interesting
difference is that the more metal-rich local Cepheids and the more metal-poor
outer disk Cepheids show differences in the sign of $\Delta \log g$. We can
only speculate that this may be an artifact of systematic differences in
the model atmospheres due to metallicity for such low-gravity stars.

The model atmospheres we have computed rely on the 
classical plane-parallel one-dimensional assumption, which may not
be an adequate representation of the real atmosphere. 
By observing main sequence stars in the open cluster M25 (which also contains
the Cepheid U Sgr), \citet{fry97} investigated the reliability of classical 
model atmospheres and whether departures from LTE 
affect an analysis in which LTE models are 
used in combination with spectroscopically
derived stellar parameters. Their [Fe/H] values for the two M25 dwarfs agreed
with their measured value for U Sgr. The derived metallicities also matched 
previous determinations 
of dwarfs in this cluster. They also followed several Cepheids throughout the
entire pulsation cycle and found essentially identical iron abundances even as 
\teff~varied by over 1000~K. Identical abundance ratios for [Si/Fe], 
[Ca/Fe], and [Ti/Fe] were also obtained in Cepheids at different phases. 
\citet{fry97} took this as evidence that a traditional spectroscopic 
analysis using classical model atmospheres can provide self-consistent 
and therefore reliable results
for Cepheids. It is worth reiterating that Andrievsky and collaborators adopted 
the \citet{fry97} spectroscopic \teff~scale in their analyses. 

Andrievsky's concerns about a traditional spectroscopic analysis are reasonable. 
However, we offer four pieces of evidence suggesting that our analyses are 
self-consistent. First, visual inspection of the spectra suggest that
HQ Per and CR Ori must have virtually identical stellar parameters 
(see Figure \ref{fig:spectra}) which is confirmed by our analysis. Similarly,
a glance at the spectra of GP Per and GV Aur suggest that their stellar
parameters and/or compositions must differ (see Figure \ref{fig:eu}). 
Once more our analysis confirms that indeed the metallicities are different. 
Second, the comparison of abundances between this study and \citet{fry97}
as well as Andrievsky et al.\ reveal that the differences are small and 
comparable to our estimates of the measurement uncertainties. Furthermore,
the abundance differences $\Delta$[X/Fe] are not correlated with \teff~though
$\Delta$[Fe/H] may be correlated with \teff. Third,
we confirm the \citet{fry97} result that abundance ratios [Fe/H] and [X/Fe]
are not a function of phase and/or \teff. In some cases, \teff~differs
by over 1000~K. Lastly, we find that our spectroscopic gravities also
produce ionization equilibrium for Ti, a species which is more susceptible
to non-LTE effects than Fe due to its lower ionization potential. 

\section{COMPARISONS WITH OTHER STUDIES OF YOUNG STARS}

\subsection{Mean abundance ratios and trends with Galactocentric distance}

For the iron abundances in our program Cepheids, 
we find a mean value [Fe/H] = $-$0.60 $\pm$ 0.05 
($\sigma$ = 0.21). For the other abundances, we find mean values of
[Mg/Fe] = 0.26 $\pm$ 0.03 ($\sigma$ = 0.13), 
[Si/Fe] = 0.32 $\pm$ 0.02 ($\sigma$ = 0.10),
[Ca/Fe] = 0.18 $\pm$ 0.02 ($\sigma$ = 0.10),
[Ti/Fe] = 0.13 $\pm$ 0.03 ($\sigma$ = 0.14),
[$\alpha$/Fe] = 0.20 $\pm$ 0.02 ($\sigma$ = 0.08),
[La/Fe] = 0.37 $\pm$ 0.03 ($\sigma$ = 0.13), and 
[Eu/Fe] = 0.37 $\pm$ 0.03 ($\sigma$ = 0.12). The program
Cepheids have a mean distance 14.3 kpc ($\sigma$ = 1.5 kpc). 
In the outer Galactic disk, our program Cepheids have subsolar ratios
of [Fe/H] along with elevated ratios of [$\alpha$/Fe], [La/Fe], and [Eu/Fe]. 
For all abundance ratios [Fe/H] and [X/Fe], we find a large dispersion
at a given Galactocentric distance. For [Fe/H], [$\alpha$/Fe], [La/Fe], and
[Eu/Fe], the dispersion exceeds the measurement
uncertainties suggesting that the outer Galactic disk is not a well-mixed 
single population. 

\begin{figure}
\epsscale{0.8}
\plotone{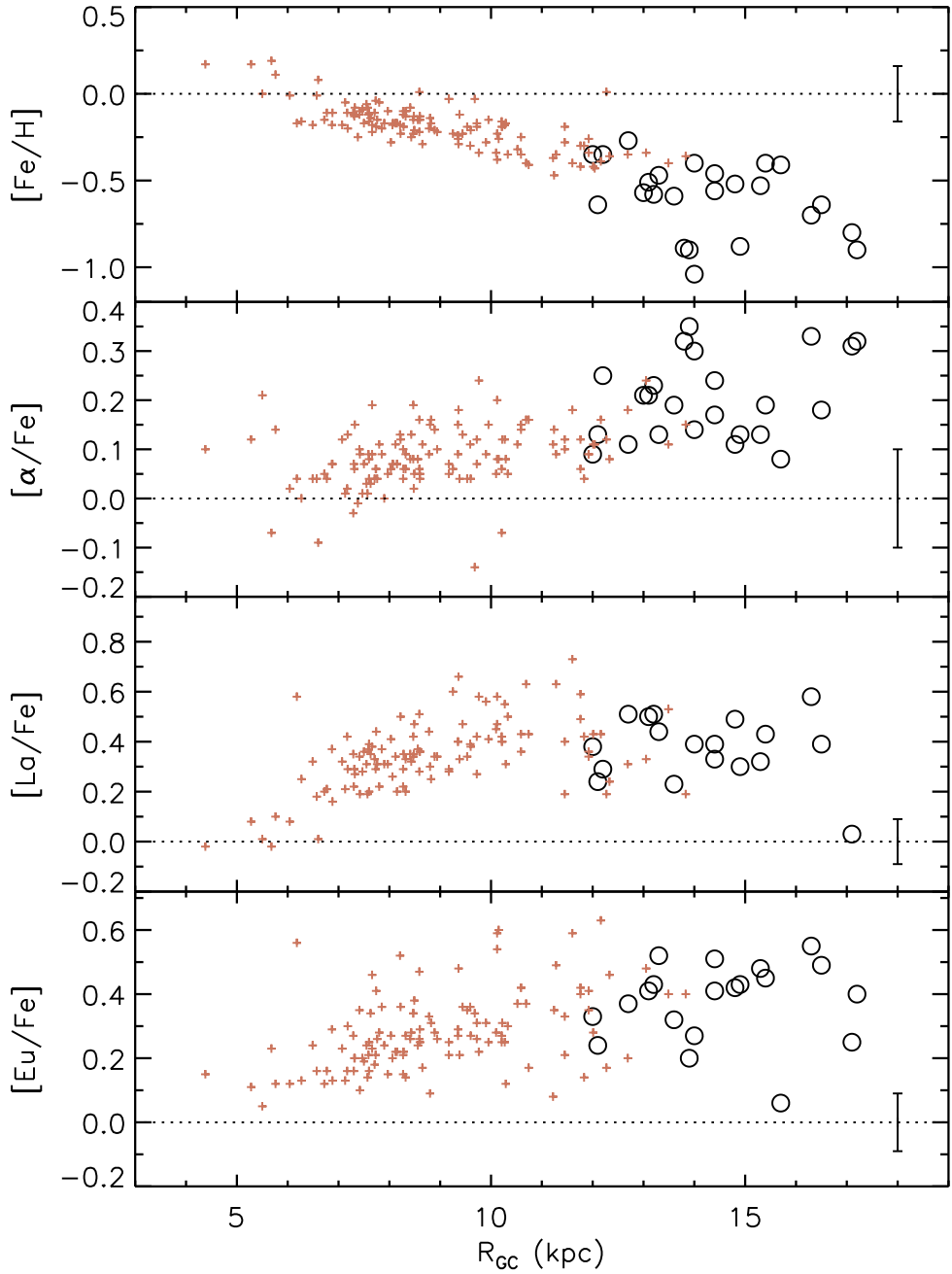}
\caption{Abundance ratios [Fe/H], [$\alpha$/Fe], [La/Fe], and [Eu/Fe] versus
Galactocentric distance \rgc~(kpc). The program Cepheids are 
represented by open black circles and
the red plus signs are Cepheids from Andrievsky and collaborators. A
representative error bar for the program Cepheids is shown. 
\label{fig:rgc1}}
\end{figure}

\begin{figure}
\epsscale{0.8}
\plotone{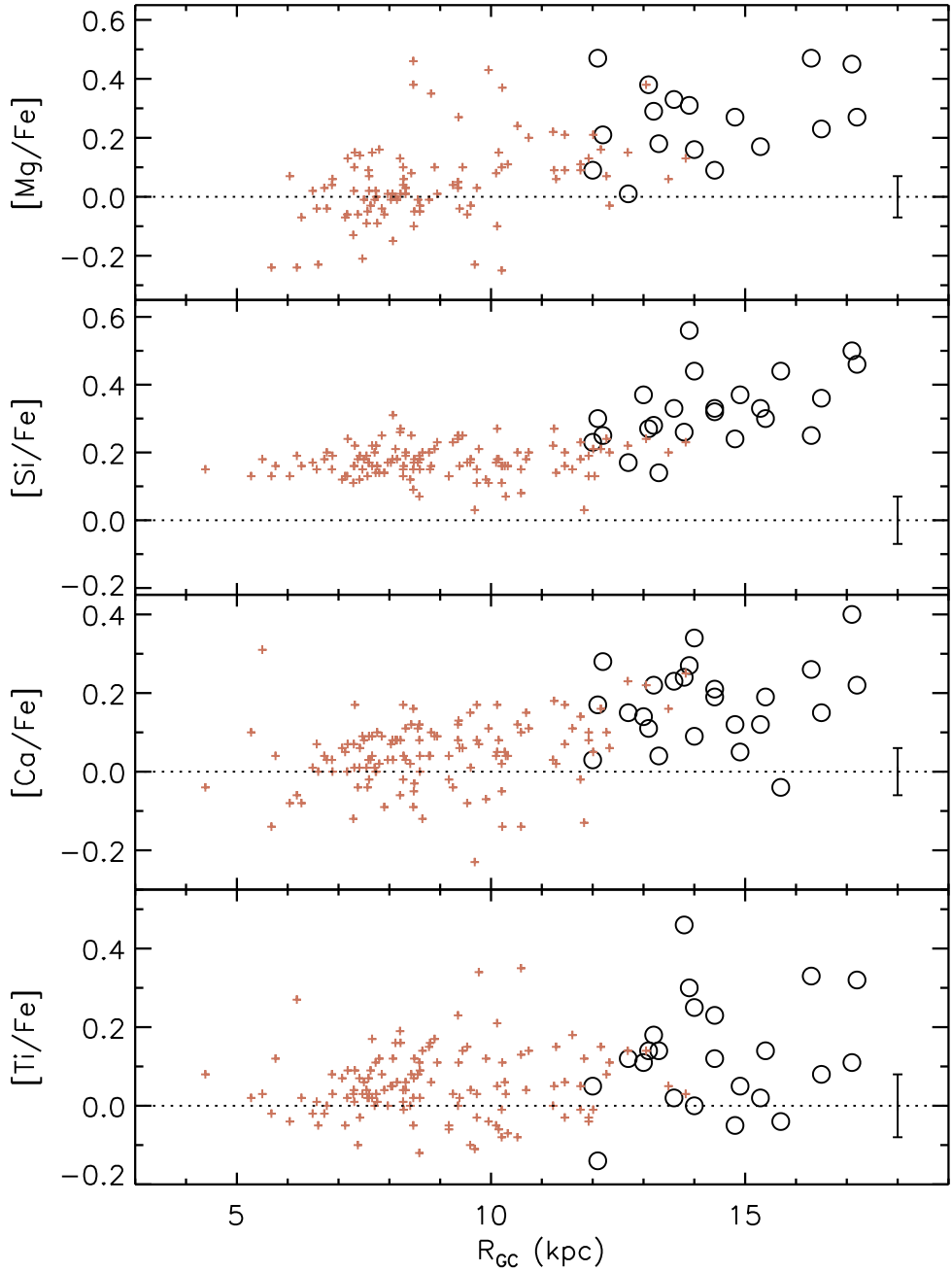}
\caption{Abundance ratios [Mg/Fe], [Si/Fe], [Ca/Fe], and [Ti/Fe] versus
Galactocentric distance \rgc~(kpc). The symbols are the same as
in Figure \ref{fig:rgc1}. A
representative error bar for the program Cepheids is shown.
\label{fig:rgc2}}
\end{figure}

In Figures \ref{fig:rgc1} and \ref{fig:rgc2}, we plot the abundance 
ratios [Fe/H] and [X/Fe] versus 
Galactocentric distance. 
By imposing a linear least squares fit to our outer disk program 
Cepheids which lie between 12 and 17.2 kpc, we find the following 
slopes and associated uncertainties: 
d[Fe/H]/d\rgc~= $-$0.052 $\pm$ 0.022 dex kpc$^{-1}$, 
d[Mg/Fe]/d\rgc~= 0.023 $\pm$ 0.010 dex kpc$^{-1}$,
d[Si/Fe]/d\rgc~= 0.033 $\pm$ 0.010 dex kpc$^{-1}$,
d[Ca/Fe]/d\rgc~= 0.012 $\pm$ 0.008 dex kpc$^{-1}$,
d[Ti/Fe]/d\rgc~= 0.019 $\pm$ 0.011 dex kpc$^{-1}$,
d[$\alpha$/Fe]/d\rgc~= 0.016 $\pm$ 0.014 dex kpc$^{-1}$,
d[La/Fe]/d\rgc~= $-$0.013 $\pm$ 0.014 dex kpc$^{-1}$, and 
d[Eu/Fe]/d\rgc~= 0.009 $\pm$ 0.013 dex kpc$^{-1}$. 
Some of these abundance gradients are statistically significant at 
the 2 or 3 $\sigma$ level. However, an inspection of Figures \ref{fig:rgc1} 
and \ref{fig:rgc2} suggest that it is not clear that a linear trend is the
appropriate function to apply, a caution we raised as well in Papers I and II. 

\subsection{Comparison with Andrievsky's Cepheid samples}

Andrievsky provided a thorough account of the Cepheid abundance trends as 
a function of Galactocentric distance out to 15 kpc. Briefly, 
he and his collaborators 
\citep{andrievsky02a,andrievsky02b,andrievsky02c,luck03,andrievsky04} 
suggested that the Galaxy can 
be divided into three zones because a single linear radial abundance gradient 
is inadequate to fit to the data.  
Zone 1 covers 4.0 $\le R_{\rm GC} \le$ 6.6, zone 2 covers 
6.6 $\le R_{\rm GC} \le$ 10.6, and zone 3 covers 10.6 $\le R_{\rm GC} \le$ 14.6.
Since a linear function is not 
supported by their data, it is more accurate to describe the global trend 
with Galactocentric distance as a ``radial abundance distribution'' even though
a particular range in \rgc~may be fitted with a linear function. 
In zone 1, the iron abundance decreases sharply with increasing Galactocentric 
distance with d[Fe/H]/d\rgc~= $-$0.128 $\pm$ 0.029 dex kpc$^{-1}$. 
In zone 2, the iron abundance decreases more gradually with 
d[Fe/H]/d\rgc~= $-$0.044 $\pm$ 0.004 dex kpc$^{-1}$. In zone 3, the 
iron abundance is essentially flat with 
d[Fe/H]/d\rgc~= $-$0.004 $\pm$ 0.011 dex kpc$^{-1}$. (Note that the
formal linear slope for our data more closely matches Andrievsky's zone 2
rather than their zone 3 though our Cepheids are located in zone 3 and beyond.) 
The mean abundance [Fe/H] for each zone was different with the mean abundance 
decreasing with increasing distance. 
Andrievsky showed that the abundance ratios [X/H] for many other elements 
behave similarly to [Fe/H] in these three zones. For almost all elements 
in each zone, the dispersion about the mean relation was low 
suggesting that the interstellar medium was well mixed at the time
the Cepheids formed, unlike what we appear to have found in the outer disk. 

\subsection{Comparison with OB stars} 

OB stars are hot and young with lifetimes and ages comparable to the young 
Cepheids. They therefore offer an independent check on both the radial 
abundance distribution as well as the dispersion in abundance ratios. 

\begin{figure}
\epsscale{0.8}
\plotone{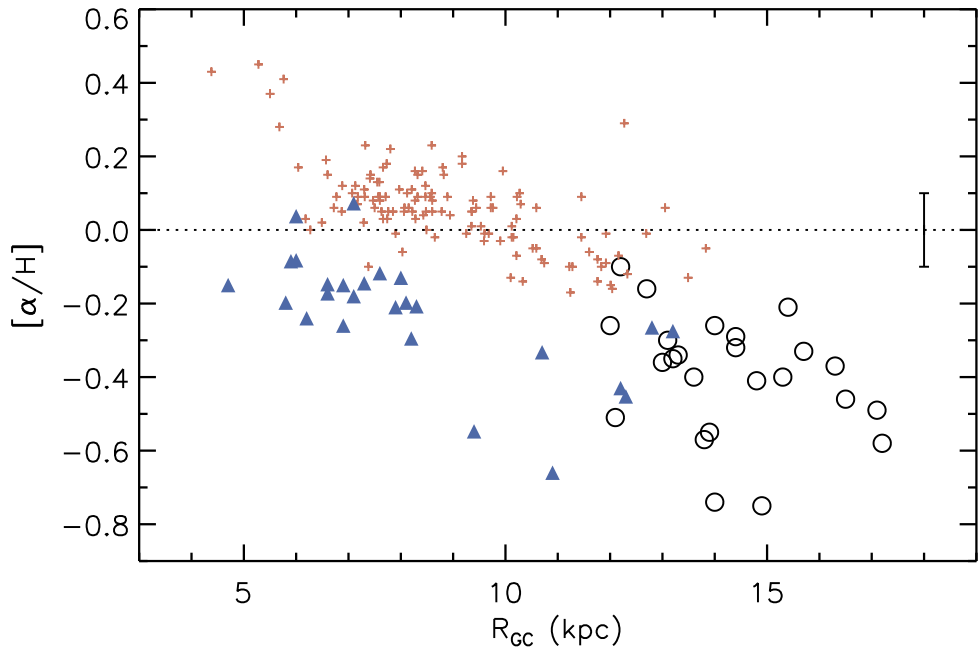}
\caption{Abundance ratio [$\alpha$/H] versus Galactocentric distance \rgc~(kpc). 
The program Cepheids are represented by open black circles, 
the red plus signs are Cepheids from Andrievsky and collaborators, and
the filled blue triangles are OB stars from \citet{daflon04}. A representative
error bar is shown.  
\label{fig:ob}}
\end{figure}

\citet[and references therein]{daflon04} measured abundances of 
C, N, O, Mg, Al, Si, and S in a 
sample of 69 young OB stars 
belonging to 25 open clusters, OB associations, and H {\sc ii} regions.
These objects spanned Galactocentric distances 4.7 $\le$ \rgc~$\le$ 13.2 
kpc. Unfortunately, Fe cannot be measured in OB stars and 
the \citet{daflon04} sample does not extend beyond
14 kpc, the regime in which the Cepheids display a 
large dispersion
in iron abundances. 
In Figure \ref{fig:ob}, we plot [$\alpha$/H] versus
Galactocentric distance for our Cepheids 
($\alpha$ = Mg+Si+Ca+Ti), Andrievsky's Cepheids 
($\alpha$ = Mg+Si+Ca+Ti), 
and the \citet{daflon04} sample of OB stars ($\alpha$ = O+Mg+Si+S). 
Prior to making a direct
comparison between the OB stars and the Cepheids, we caution that
independent analysis techniques are employed for the analysis of
these different objects. There is a offset between the Cepheids
and OB stars with the OB stars showing lower [$\alpha$/H] by roughly 0.3 
dex at a given Galactocentric distance. It is not clear if this 
offset is real or whether it reflects the systematic differences 
between the analysis techniques. For example, the much hotter OB stars 
are more susceptible to non-LTE effects. If this
offset is real, then it would be extremely difficult to explain given the 
comparable ages and lifetimes of the OB stars and Cepheids. For the 
range of distances spanned by the OB stars, the radial abundance gradient 
appears rather similar between the OB stars and Cepheids. Interestingly, 
the most distant OB stars  
also appear to exhibit a dispersion in 
abundances. The amplitude of the dispersion is about 0.5 dex and 
is seen in O, Mg, and Si as well as $\alpha$. The differences 
observed are not due to lower S/N spectra which ranged from 70 to 300. 

We take two main results from the OB stars. 
Firstly, the abundances tend to decrease with
increasing Galactocentric radius. In the range of distances covered
by both the OB stars and Cepheids, the radial abundance distributions appear
similar. Secondly, there is a dispersion
in abundances at large distances. While there is also a scatter at smaller
distances, the scatter in distant OB stars is greater than for the local
OB stars. In particular, we note that the maximum
dispersion is for [Si/H] and is roughly 0.8 dex, a value comparable
to the amplitude of the dispersion seen in [Fe/H] in our Cepheids. 
Both points illustrate that the young OB stars and the young Cepheids 
likely have similar mean abundances, radial abundance distributions, 
and a dispersion in abundance ratios at a given Galactocentric distance. 
While the dispersion in abundance ratios may be a natural consequence of low 
densities in the outer Galactic disk, there are other possible explanations. 

\section{DISCUSSION}

\subsection{Introduction}

\begin{figure}
\epsscale{0.8}
\plotone{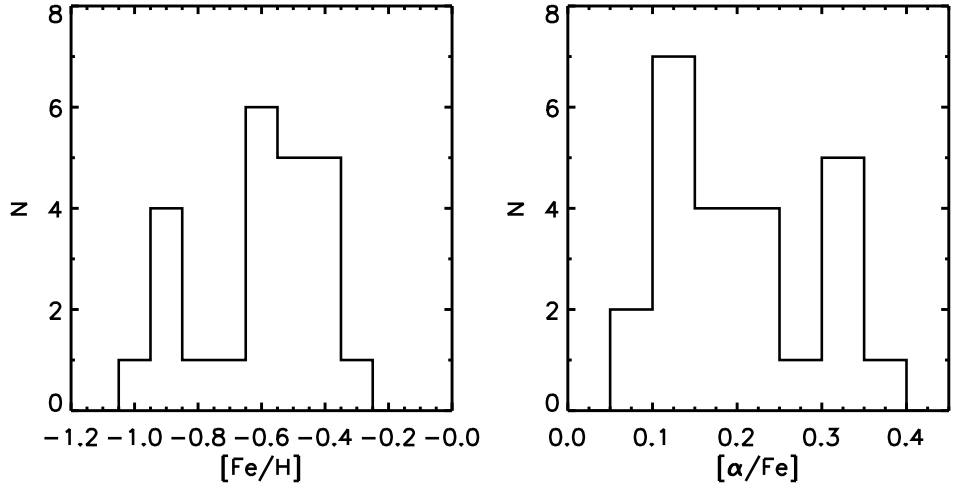}
\caption{Metallicity distribution functions for [Fe/H] and [$\alpha$/Fe]. 
\label{fig:hist}}
\end{figure}

In Figure \ref{fig:rgc1}, we show [Fe/H] and [$\alpha$/Fe] versus
Galactocentric distance for our sample of outer disk Cepheids. In this
Figure we also include Andrievsky's sample 
noting that the abundance ratios have
been shifted onto our system after an abundance comparison of a 
common subsample. (Shifts of $-$0.16, 0.11, 0.12, and 0.19 were applied to
Andrievsky's [Fe/H], [$\alpha$/Fe], [La/Fe], and [Eu/Fe] respectively.)
For [Fe/H] and [$\alpha$/Fe], our data appear
to continue the trends with \rgc~seen by Andrievsky. However, beyond 
\rgc~= 14 kpc, a subset of the sample have unusually 
low [Fe/H] as well as unusually high 
[$\alpha$/Fe]. There is a hint that in the outer disk the iron abundances may 
have a bimodal distribution with peaks at [Fe/H] = $-$0.5 and [Fe/H] = $-$0.9 
(see Figure \ref{fig:hist}). In the same Figure, the abundance ratio 
[$\alpha$/Fe] may also exhibit a bimodal distribution. 
(Both distributions are somewhat sensitive to the binning. If the bin 
centers are shifted by half a bin width, [Fe/H] still exhibits 
bimodality while [$\alpha$/Fe] instead shows an asymmetric distribution.)
We offer two explanations
for this apparent bimodality. 
The first possibility is that the outer disk Cepheids represent a single 
population which has been subject to inhomogeneous chemical 
evolution. Such an idea is consistent with the fact that for 
[Fe/H], [$\alpha$/Fe], [La/Fe], and [Eu/Fe], the dispersion exceeds the 
measurement uncertainties. Inhomogeneous chemical evolution may be a 
reasonable expectation in the low density outer Galactic disk. 
The second possibility is that the outer disk
Cepheids represent two (or more) separate populations whose star formation 
histories and nucleosynthetic histories are distinct despite their young 
ages. This suggests one or more merger events may be underway in the outer 
disk. Such an idea is
compatible with the hierarchical assembly of galaxies predicted from 
``$\Lambda$CDM'' cosmological simulations that successfully reproduce 
the observed large scale structure of the Universe. 

\subsection{Do the Cepheids represent a single population?}

\subsubsection{Expectations} 

Before exploring the possibility that the Cepheids represent a 
single population, we offer a brief discussion of 
plausible expectations for the chemical abundances in the outer disk. 

If the outer disk has been undergoing star formation since
the Galaxy formed, and has evolved essentially in isolation,
with negligible amounts of infall or mergers, we expect to 
find the following properties.
A. The outer disk
Cepheids should be more metal-poor than the inner disk Cepheids, due
to the slower pace of chemical evolution in these lower density regions.
Numerous chemical evolution models (e.g., \citealt{hou00,alc01,chiappini01})
support this
basic idea. 
 B. The Cepheids should have essentially solar values of
     [$\alpha$/Fe]. The primary factor involved in the behavior
     of [$\alpha$/Fe] vs.\ [Fe/H] is the star formation rate
     and the time since it began. Assuming durations of star
     formation in the inner and outer disk, the slower star
     formation rate in the outer disk would have led to the
     appearance of SNe~Ia ejecta when the [Fe/H] was still
     quite low. Therefore, [$\alpha$/Fe] would have reached
     solar values after a comparable amount of time but at
     a lower [Fe/H]. If star formation in the outer disk
     began very recently, so that the contributions of SNe~Ia
     had not had time to contribute significantly, only then
     would we anticipate [$\alpha$/Fe] values substantially
     above solar, and the ages of the oldest outer disk clusters
     are comparable to the ages of the oldest inner disk 
     open clusters.
C. We might expect a range in
[Fe/H] for the outer disk Cepheids, and among field stars and clusters
of all ages, simply because of the low densities. Recall that
[Fe/H] is affected by the mean distance of star-forming regions from
the sources of nucleosynthesis products, but that [$\alpha$/Fe] should
not be so affected. D. Despite a slower pace of star formation in the
outer disk compared to the solar regions, the steady progress of
chemical enrichment should lead to somewhat higher mean metallicities
as a function of time. The Cepheids should be at least as 
metal-rich as the older field red giants and old open clusters.
We explore now how these predictions are or are not satisfied.

Andrievsky's sample of Cepheids already demonstrated that the inner
disk Cepheids are more metal-rich than the outer disk Cepheids. 
In Figure \ref{fig:rgc1}, our distant Cepheids are considerably more
metal-poor than the inner disk Cepheids. Therefore, the first expectation
is satisfied. 

In considering the second expectation, that the young Cepheids should
have [$\alpha$/Fe] $\approx$ 0 despite their low iron abundances, we again
rely initially upon Andrievsky's results. His data showed that the
outer disk Cepheids may have slight enhancements in [$\alpha$/Fe] 
relative to the inner disk Cepheids. In Figure \ref{fig:rgc1}, our
distant Cepheids have considerably higher [$\alpha$/Fe] ratios than the 
inner disk Cepheids. Therefore, the second expectation is not confirmed
by the data. 

The third expectation is that the outer disk Cepheids should have a range
in [Fe/H] due to the lower densities. Andrievsky's results showed that
such a dispersion was not evident. For our sample of more distant Cepheids, 
we do find a large range in [Fe/H] as expected. However, we now need
to consider whether or not the outer disk open clusters and field giants
exhibit a range in [Fe/H]. A comparison between Cepheids, field
stars, and open clusters is also required to assess 
the fourth expectation. 

\subsubsection{A comparison with field giants and old open clusters}

In the limited samples of outer disk open clusters and field giants
presented in Papers I and II,
the abundance ratio [Fe/H] appeared to reach a basement value with 
very little dispersion about the mean value, [Fe/H] $\approx$ $-$0.5, 
beyond 12 kpc. Similarly, the abundance ratio [$\alpha$/Fe] appeared
to reach a ceiling with little dispersion about the mean value, 
[$\alpha$/Fe] $\approx$ 0.2, beyond 12 kpc. 

\begin{figure}
\epsscale{0.8}
\plotone{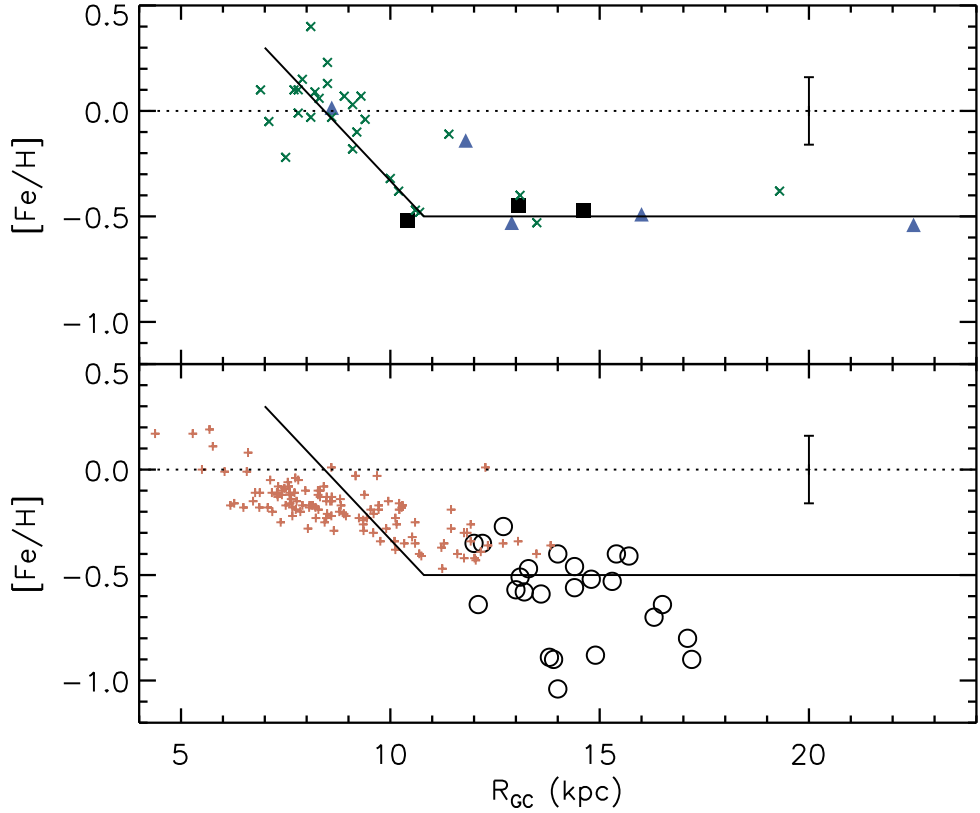}
\caption{Abundance ratio [Fe/H] versus Galactocentric distance \rgc~(kpc). 
In the upper panel, we plot the open cluster giants from Paper I (filled 
blue triangles), the field giants from Paper II (filled black squares), 
and open clusters (green crosses) from the compilation by \citet{friel05}. 
In the lower panel, we plot the program Cepheids (open black circles) and
the Cepheids from Andrievsky and collaborators (red plus signs).  
In both panels, a representative error bar is shown. A schematic 
line fit to the data in the upper panel is shown in both panels. 
\label{fig:rgc.fe}}
\end{figure}

In Figure \ref{fig:rgc.fe} we plot [Fe/H] versus 
Galactocentric distance for the Cepheids, open clusters, and field
stars. A schematic line illustrates the behavior of the open clusters
and field stars. We impose this line upon the results from the 
Cepheids. Within the limits
of the available data, the Cepheids show more scatter in [Fe/H] 
in the outer disk (\rgc~$>$ 10 kpc) than do the open clusters
and field giants. Therefore, the dispersion 
in [Fe/H] seen in the outer disk Cepheids 
does not appear to be a feature of the 
open clusters and field giants and the third expectation is not 
completely satisfied.

The schematic line shown in Figure \ref{fig:rgc.fe} also shows that
the Cepheids reach significantly lower metallicities than do the
open clusters and field stars at a given Galactocentric distance. 
As just described, in a ``closed box'' model for the evolution of the Galaxy, 
at the same Galactocentric distances, the metallicities of the young
Cepheids must always be greater than or equal to those of the older open 
clusters and field stars. Therefore, the fourth expectation is not 
supported by the data. The failure of the data to meet this expectation
offers the strongest evidence
that a ``closed box'' model for the evolution of our Galaxy is inappropriate 
and that the compositions of the outer disk open clusters, field giants, and 
Cepheids cannot be explained as a simple evolutionary process of an isolated
ensemble of stars and gas. The second and third expectations were also 
not satisfied. 

In Figure \ref{fig:rgc.fe}, there appears to be a 0.2-0.3 dex difference  
between the iron abundances in the open clusters and the Andrievsky et al.\ 
Cepheids in the range centered on 8 kpc. We can attribute 0.16 dex to the
offset applied to the Andrievsky Cepheids as discussed in 
Section \ref{sec:luckcomp}. It is not clear if the remaining difference
is real, perhaps as a consequence of differing ages, or a result of the 
analysis techniques.

\begin{figure}
\epsscale{0.8}
\plotone{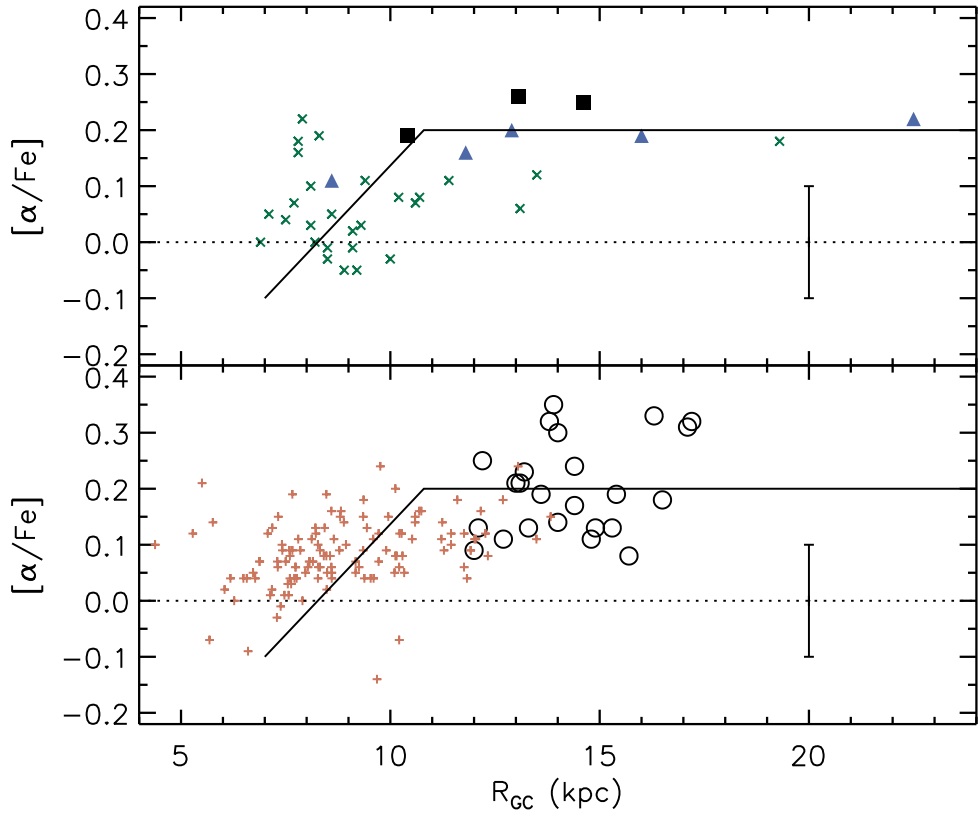}
\caption{Same as Figure \ref{fig:rgc.fe} but for the abundance ratio 
[$\alpha$/Fe]. 
\label{fig:rgc.al}}
\end{figure}

We now compare the ratio [$\alpha$/Fe] versus Galactocentric distance for 
the Cepheids, open clusters, and field stars. 
In Figure \ref{fig:rgc.al}, we again show a schematic line to highlight the 
behavior of the open clusters and field stars and impose this line onto the 
Cepheids. We note that within the available data, the Cepheids show a 
greater scatter in [$\alpha$/Fe] in the outer disk than do the open clusters
and field stars. Further, a number of Cepheids reach 
higher [$\alpha$/Fe] than do the open clusters and field stars. Elevated
ratios of [$\alpha$/Fe] reveal that recent star formation has occurred 
with Type II supernovae contributing ejecta to the interstellar gas from
which the Cepheids formed. These overabundances of [$\alpha$/Fe] also 
suggest that Type Ia supernovae did not contribute as significantly to the
outer disk's chemical evolution as is the case for local Cepheids. 

For the open clusters and field stars in the outer disk, the 
$r$-process element Eu showed enhanced ratios [Eu/Fe] = 0.3 to 0.5. 
Since the $r$-process is believed to occur in massive stars, we interpreted 
this as confirmation that recent star formation had taken place in the 
outer disk when the field stars and clusters formed several Gyrs ago, 
as inferred from high [$\alpha$/Fe]. 
The younger outer disk Cepheids also show enhanced ratios of [Eu/Fe] 
and [$\alpha$/Fe] compared to the solar value. Once again we attribute 
the enhancements of elements synthesized in massive stars to even more 
recent star formation in the outer disk.

\subsection{Do the Cepheids represent separate populations?} 

\begin{figure}
\epsscale{0.8}
\plotone{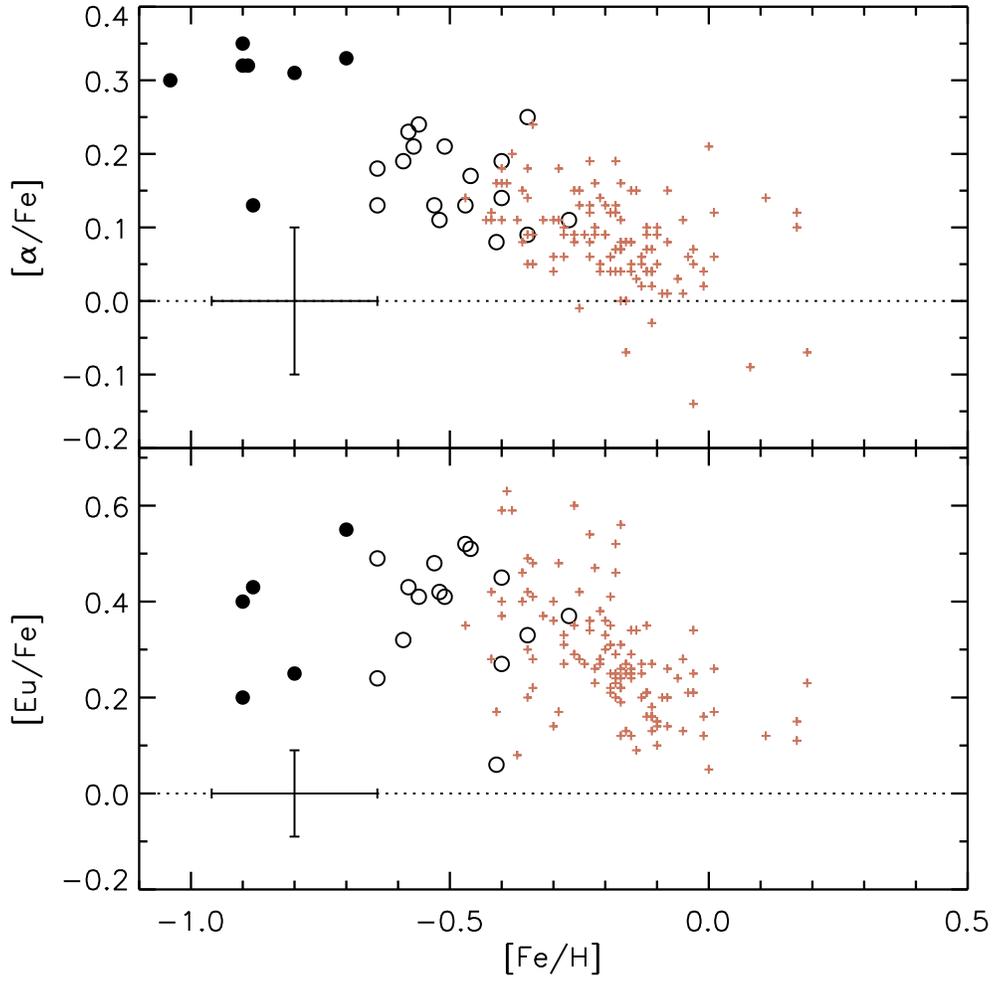}
\caption{Abundance ratios [$\alpha$/Fe] and [Eu/Fe] versus [Fe/H]. 
Open black circles are our ``Galactic Cepheids'', closed black circles 
are our ``Merger Cepheids'', and red plus signs are Andrievsky's Cepheids.
A representative error bar for the program Cepheids is shown.
\label{fig:x2fe}}
\end{figure}

We consider now one additional dimension of complexity 
which may help explain our results. A simple model for the
bimodality in [Fe/H] and [$\alpha$/Fe] 
seen in Figure \ref{fig:hist} is that the
Cepheids represent different stellar populations. 
In Figure \ref{fig:rgc.fe}, we see that a modest majority of our 
program Cepheids
appear to reflect a smooth transition to the inner disk Cepheids
studied by Andrievsky, and we adopt a model that describes those
Cepheids as having been formed within the Galactic interstellar
medium. We refer to these stars as the ``Galactic Cepheids".
For the most metal-poor and distant Cepheids with 
$R_{GC}$ $>$ 13.5 kpc and [Fe/H] $\leq\ -0.65$, we speculate 
that some stars may have formed under the influence of a significant 
on-going merger or accretion event. We refer to these
stars as the ``Merger Cepheids''. One signature of such a 
merger event would be a significant difference in the chemical abundance
ratios between the ``Galactic Cepheids'' and the ``Merger Cepheids'', 
which, recall, are independent of the distance from the sites
of nucleosynthesis. Figure \ref{fig:x2fe} shows that this suggested 
sub-sample of our program stars (the ``Merger Cepheids'') are very unusual. 
Not only are the ``Merger Cepheids'' more metal-poor
than the larger subsample of ``Galactic Cepheids", but their 
[$\alpha$/Fe] ratios are also markedly different. The differences in
the ages of Cepheids is unlikely to be even comparable to, much
less longer than, the timescale for the appearance of SNe~Ia
in a rich star-forming environment, so these differences should
reflect a major difference in the origins of the gas out of which
the two sets of Cepheids formed. It seems most likely that these
Cepheids are related, somehow, to the on-going merger events discussed
below.

While we have described the majority of our program stars as having been
formed as part of the normal evolution of the Galactic disk, we
stress that Figures \ref{fig:rgc.fe} and \ref{fig:rgc.al} suggest that 
merger and/or accretion events have been involved in the past as 
well. Consider Figure \ref{fig:rgc.fe}. Note that for the ``Galactic
Cepheids", the upper envelope of the iron abundances appears to match
the basement metallicity seen in the open clusters and field stars. 
However, the Cepheids reach this value at larger Galactocentric distances 
than do the older clusters and field stars. 
If the absence of the gradient arises 
due to a succession of accretion or merger events, as discussed by
\citet{twarog97} and \citet{warp}, the larger Galactocentric distance
of that basement iron abundance 
for the younger Cepheids may be explained most readily by
a growth in the Galactic disk. At the ages of the older clusters, several
billion years, the edge of the Galactic disk was at 10-11 kpc, while
now it is perhaps at 14 kpc or so. 

We find this signature of a succession of accretion/merger events may
provide an explanation for the unusual metallicities and abundance
patterns we have found among the Cepheids and the older open clusters.
The data suggest that the Galactic disk has grown with time, and that
signs of an on-going merger event may be seen in the chemical abundance
patterns of a subsample of young stars in the outer Galactic disk.

\subsection{Cepheids as possible probes of rapid chemical evolution}

Lower luminosity Cepheids evolve from lower luminosity main sequence stars 
whose lifetimes exceed those of higher luminosity main sequence stars. 
Therefore, higher and lower luminosity Cepheids may
probe slightly different epochs of chemical enrichment with the lower 
luminosity Cepheids extending back to more distant times. Since luminosity 
and period are famously related in Cepheids, we 
can compare abundance ratios [Fe/H] and 
[X/Fe] versus period to investigate whether or not chemical evolution has
occurred within the short timescales spanned by the Cepheids. 

\begin{figure}
\epsscale{0.8}
\plotone{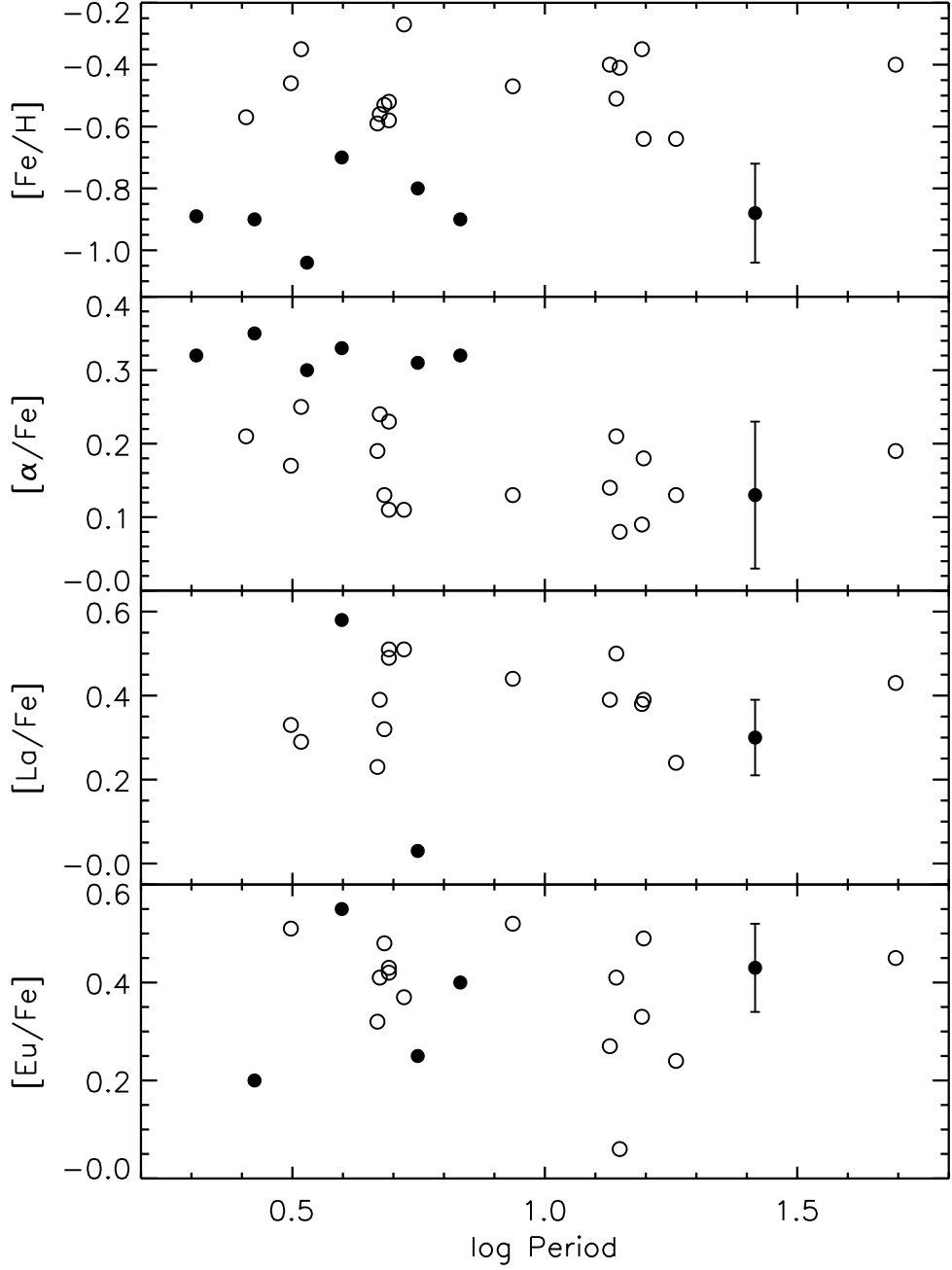}
\caption{Abundance ratios [Fe/H], [$\alpha$/Fe], [La/Fe], and
[Eu/Fe] versus $\log$ Period. Open black circles are our 
``Galactic Cepheids'' and closed black circles are our ``Merger Cepheids''.
A representative error bar is shown. 
\label{fig:alogp}}
\end{figure}

In Figure \ref{fig:alogp}, we plot [Fe/H] and [$\alpha$/Fe] versus period. 
We use different symbols to distinguish the ``Galactic Cepheids'' from the
``Merger Cepheids''. There is a hint that the ``older'' Cepheids 
(lower luminosity and therefore shorter periods) may have lower [Fe/H] and
higher [$\alpha$/Fe]. While our sample sizes are small, a trend may be
evident within both the ``Galactic'' and ``Merger'' Cepheids. 
This suggests a greater contribution from Type II 
supernovae at earlier times and that the rate of chemical evolution 
is rapid. This also suggests that chemical evolution may
have taken place even within the small time-frame spanned by the Cepheids. 
\citet{bono05} provide a period-age relation for Cepheids. For the metal-poor
shorter period Cepheids, the lifetimes are roughly 320 million years 
(assuming $Z$ = 0.004 and log P = 0.6). For the metal-rich longer period
Cepheids, the lifetimes are roughly 250 million years (assuming $Z$ = 0.01 
and log P = 1.2). So the difference in ages could be as high as 70 million 
years for the Cepheids. 
The enhanced [$\alpha$/Fe] and [Eu/Fe] already signify 
recent star formation and the possible trend between [$\alpha$/Fe] and 
period appear to confirm this finding.
In the same Figure, we plot [La/Fe] and 
[Eu/Fe] though we note that the number of lines used to derive these
elements is smaller than for Fe and $\alpha$ and therefore these
abundances are subject to greater uncertainties. La and Eu do not exhibit 
any trends with period. 

\subsection{Additional clues for chemical evolution: Neutron-capture elements}

That the younger Cepheids have lower metallicities than the older open
clusters and field stars suggests that the outer disk cannot be
described as a simple evolution of a ensemble of stars and gas. 
The possibly bimodal distribution for [Fe/H] and [$\alpha$/Fe] displayed by 
the Cepheids may reflect the fact that the Cepheids are composed of  
different populations. 
We now turn to the neutron-capture elements La and Eu to see what 
information they provide regarding chemical evolution of the Cepheids and
the merger hypothesis. 

The $s$-process element La is believed to be synthesized primarily within
low-mass AGB stars (e.g., see \citealt{busso99}
for a review). The elevated ratios of [La/Fe] in the outer
disk show that the interstellar gas from which the outer disk
Cepheids formed had been polluted by AGB stars, a result previously seen
in the open clusters and field stars. We find that the ``Merger Cepheids''
and the ``Galactic Cepheids'' have similar ratios for [La/Fe], 
0.30 and 0.39 respectively. 

We have already noted that the $r$-process element Eu shows enhancements 
in the outer disk. Eu and the $\alpha$ elements are believed to be
synthesized within massive stars and the ``Merger Cepheids'' and 
``Galactic Cepheids'' showed different values for [$\alpha$/Fe]. 
Interestingly, we note that the [Eu/Fe] ratios between the two Cepheid
populations do not differ, 0.38 for the ``Merger Cepheids'' and 
0.37 for the ``Galactic Cepheids''. This lack of
difference may be due to the fact that Eu abundances are derived from 
a single line and are therefore more uncertain than the $\alpha$ 
abundances which are derived from numerous lines. 

Both La and Eu exhibit considerable dispersions (as do Fe and $\alpha$). 
In Papers I
and II, we noted that the ratio [La/Eu], $s$-process to $r$-process material,
also displayed a scatter in the outer disk field stars and open clusters with 
no objects displaying a scaled-solar pure $r$-process or pure $s$-process
ratio. The Cepheids have a ratio of [La/Eu] that is centered at the solar 
value and exhibits a dispersion with no Cepheid having a
scaled-solar pure $r$-process or pure $s$-process distribution.
Elevated ratios of La and Eu were previously seen in Papers I and II
and similar conclusions were drawn regarding the recent star formation
and contribution of AGB stars to the chemical evolution of the outer disk. 
When we consider the ``Merger Cepheids'' and the ``Galactic Cepheids''
separately, we find that the ratios of [La/Eu] are virtually identical 
within the uncertainties, $-$0.06 and $-$0.01 respectively. 

\section{CONCLUSIONS AND FUTURE WORK}

In the previous two papers in this series, we presented the chemical
compositions for old open cluster giants and field giants in the 
outer Galactic disk. In this paper we conduct an abundance analysis 
of 24 young Cepheids located at large Galactocentric distances. 
The program Cepheids therefore allow us to study the time evolution
of the Galactic radial abundance distribution as well as the
current radial abundance distribution. 

The short lifetimes
of Cepheids ensures that their compositions reflect the recent
state of the ISM. In general, the 
abundances measured in our program Cepheids 
continue the trends with Galactocentric distance seen in the large
sample of Cepheids analyzed by Andrievsky. We also find enhancements for
[La/Fe] and [Eu/Fe] in the outer disk. For all elements, we find 
a scatter that exceeds the measurement uncertainties. The enhancements
in [Eu/Fe] and [$\alpha$/Fe] suggest that Type II supernovae have
played a greater role in the chemical evolution than have Type Ia supernovae
in the outer disk, compared to the inner disk.
The short lifetimes of the Cepheids demonstrate that recent star formation
has taken place in the outer disk. The high ratios of [La/Fe] suggest that AGB
stars
have also played a role in the evolution of the outer disk. We 
find that the ratio [La/Eu] is centered at the solar value with no
star showing a scaled-solar pure $s$-process or $r$-process value. 

The sample of Cepheids in the outer disk, although numbering only 24 stars,
has provided some tantalizing clues to the evolution of our Galaxy.
As expected, the outer disk Cepheids are on average more metal-poor than
Cepheids with smaller Galactocentric distances. However, the outer disk
Cepheids show higher abundances of the $\alpha$ elements, despite having
very similar ages to Cepheids in the inner disk. The simplest conclusion to
be drawn is that recent star formation in the outer disk has provided
a recent enhanced production of Type II supernovae relative to Type Ia
supernovae. Either the inner disk is undergoing a slower pace 
of chemical evolution, or there is enhanced star formation underway in
the outer disk. The question would then be what is causing this phenomenon
in such a low density environment? 

There are two hints in the data that point toward accretion as being the
underlying cause of the enhanced [$\alpha$/Fe] ratios. First, the
most distant Cepheids appear to have bimodal distributions of [Fe/H] 
and [$\alpha$/Fe]. The more metal-rich Cepheids of the
outer disk appear to share a similar [$\alpha$/Fe] vs.\ [Fe/H] trend
as the inner disk Cepheids. But the outlying more metal-poor outer disk
Cepheids show even higher [$\alpha$/Fe] values, suggesting a
separate history. Among these stars, it is intriguing that the most
recently formed such star (with the longest period) has a significantly
lower [$\alpha$/Fe] values than the four shorter period (older) Cepheids,
hinting that the sources of nucleosynthesis may have been changed on
a timescale of only a few tens of millions of years. Alternately, if
accretion is responsible for the difference, it may be that the four
older stars formed from gas that was richer in material from a
merging galaxy while the more recently formed star emerged from
gas more throughly mixed with Galactic gas.

A comparison between the young Cepheids and older field stars
and open cluster in the outer disk also provides clues about the evolution
of the outer Galactic disk. Disregarding the apparent bimodality of
the outer disk Cepheids, one must confront the observation that the
younger Cepheids are more metal-poor than the older clusters
and field stars. Accepting the bimodality, and concentrating on only
that Cepheids that appear to continue the trend of [$\alpha$/Fe] vs.\ [Fe/H]
seen in inner disk Cepheids, one finds a behavior similar to the
older clusters and field stars. [Fe/H] appears to reach a basement
value of about $-0.5$ while [$\alpha$/Fe] reaches a ceiling of
+0.15. However, the Cepheids reach such values only for $R_{\rm GC} > 14$ kpc,
while the older clusters and stars do so at 10 to 11 kpc. This suggests
that the outer {\em stellar} disk has grown in radius by several kpc
in the past several billion years. This suggests a past history of
accretion, in addition to that apparently recently underway.

Accretion events have been a common theme in other recent studies of the
Galaxy's outer disk.  
The Sloan Digital Sky Survey has detected the ``Monoceros
Ring" \citep{newberg02}. \citet{yanny03} have found
clear dynamical evidence for a structure at $\ell$ = 198,
$b = -27$ that may be part of a more extensive merger remnant.
The 
2MASS database and its ability to distinguish M dwarfs from 
M giants has led to the identification of the candidate ``Canis Major galaxy",
whose
center lies near $\ell\ \approx\ 244$, $b \approx\ -8$ 
\citep{martin04,martin05,bellazzini04,delgado05}. 
Radial 
velocities of the photometrically identified streams has confirmed 
the existence of unique streams \citep{martin05,penarrubia05,conn05}. 
Are any of these related to the
unusual star formation history we are suggesting for the outer disk?
Certainly all the evidence points to on-going accretion episodes,
but how may we distinguish individual events?

The referee has drawn our attention to the fact that the metal-poor
     but $\alpha$-rich Cepheids are not distributed as widely as the
     other Cepheids in Table~1. These seven stars, CI~Per, EW~Aur,
     FO~Cas, GP~Per, IO~Cas, NY~Cas, and OT~Per, all lie at low
     Galactic latitudes (and six between $b = -2$ and $-4$) and between
     $\ell =$ 119 and 166 degrees. Could these stars' locations be
     a clue about a possible merger origin? If there is a relation
     to an already-suggested merger event, its location is more
     consistent with Canis Major than with the Monoceros Ring.
     But even so, we are reluctant to ascribe much significance
     to the spatial locations of these ``Merger Cepheids". They are,
     after all, very spread out in longitude. Further, it is not clear
     how uniform are the searches for distant Cepheids. More extensive
     work in even one hemisphere than another could create such
     an apparent grouping (and all these Cepheids are northern ones).
     A more thorough search for Cepheids and, especially, the more
     abundant but comparably young OB stars may be needed to resolve
     this question. 

We argue that detailed chemical composition 
studies such as we have undertaken here and in
Papers I and II are critical to understanding the complex 
history of our Galaxy. Distinct chemical evolution patterns
are as important, if not more so, in probing the relationship between
candidate streams and merging galaxies to one another and to the
stars in the outer Galactic disk. Unfortunately, little work has
been done as yet.
So far, only 3 stars in Canis Major have been analyzed
\citep{sbordone05}, and the data employed had 
S/N=40 per pixel at 5800\AA~which is less than ideal (in our opinion). 
But the results are intriguing.
The abundance ratios found by \citet{sbordone05} do not appear to match
those seen in our samples of outer disk stars. Specifically, the Canis
Major candidates have low [$\alpha$/Fe] while our outer disk stars all
show enhancements in [$\alpha$/Fe]. Low abundances of [$\alpha$/Fe] 
are a signature of the current dwarf spheroidals orbiting our Galaxy 
\citep{venn04}. Enhancements in [$\alpha$/Fe] in the ``Merger Cepheids''
may therefore suggest that these Cepheids formed as a result of star
formation triggered by the merger rather than forming within the dwarf 
galaxy. 
Unfortunately, detailed stellar abundance ratios are unable to 
offer a clearer picture of the exact mechanism. While we speculate 
that the star formation was triggered by merging gas, we cannot say 
whether that gas was pristine or pre-enriched.
Detailed abundance ratios in a large
sample of candidate members of the Monoceros ring and/or Canis Major Galaxy
are required. The measured abundance ratios would confirm if the outer disk
has been growing via the merger of dwarf galaxies as well as providing
the chemical history of these small galaxies. Such a result 
would have profound implications not only for our understanding of the
evolution of our Galaxy but also for $\Lambda$CDM cosmology. 

\acknowledgments

We thank the anonymous referee for many helpful suggestions and comments. 
We are extremely grateful to the National Science Foundation for
their financial support through grants grants AST 96-19381, AST 99-88156, 
and AST 03-05431 to the University of North Carolina. 

\clearpage

\begin{deluxetable}{lccccrrrrcccrcc} 
\tabletypesize{\scriptsize}
\rotate
\tablecolumns{13} 
\tablewidth{0pc} 
\tablecaption{Program stars.\label{tab:basic}}
\tablehead{ 
\colhead{Name} & 
\colhead{Source\tablenotemark{a}} &
\colhead{RA} & 
\colhead{Dec} & 
\colhead{$l$} & 
\colhead{$b$} & 
\colhead{Period} & 
\colhead{Exposure} &
\colhead{S/N\tablenotemark{b}} &
\colhead{HJD} & 
\colhead{Phase} & 
\colhead{$<V>$} & 
\colhead{$<K>$} & 
\colhead{E(B-V)} &
\colhead{$R_{\rm GC}$} \\

\colhead{} & 
\colhead{} &
\colhead{J2000.0} & 
\colhead{J2000.0} & 
\colhead{(deg)} & 
\colhead{(deg)} & 
\colhead{d} & 
\colhead{Time (s)} &
\colhead{} &
\colhead{$-$2,450,000} & 
\colhead{} & 
\colhead{(mag)} & 
\colhead{CIT} & 
\colhead{} &
\colhead{kpc} 
}
\startdata
CE Pup & C & 08 14 08.0 & $-$42 34 05 & 259.2 & $-$4.4 & 49.53 & 6000 & 77 & 1206.8201 & 0.434 & 11.96 & 7.533 & 0.74 & 15.4 \\
CI Per & D & 02 05 02.3 & +57 08 35 & 132.8 & $-$4.3 & 3.38 & 7200 & 85 & 0811.6732 & 0.963 & 12.68 & 10.332 & 0.28 & 14.0 \\
CR Ori & M & 06 05 44.9 & +13 14 23 & 195.9 & $-$3.9 & 4.91 & 3600 & 60 & 1206.5626 & 0.687 & 12.30 & 9.195 & 0.56 & 13.2 \\
CU Mon & D & 06 32 46.8 & +00 02 35 & 210.8 & $-$4.1 & 4.71 & 7200 & 85 & 1205.5528 & 0.809 & 13.63 & 9.894 & 0.79 & 14.4 \\
CY Aur & D & 04 57 40.1 & +46 05 33 & 160.5 & 2.0 & 13.85 & 1800 & 42 & 0809.8810 & 0.128 & 11.89 & 7.716 & 0.81 & 13.1 \\
EE Mon & M & 06 50 48.7 & $-$07 58 50 & 220.0 & $-$3.8 & 4.81 & 7200 & 85 & 0832.7457 & 0.467 & 12.50 & 10.141 & 0.49 & 15.3 \\
ER Aur & D & 05 13 10.0 & +41 59 26 & 165.5 & 1.7 & 15.69 & 1800 & 42 & 0809.7541 & 0.859 & 11.53 & 8.480 & 0.52 & 16.5 \\
EW Aur & D & 04 51 24.9 & +38 11 19 & 166.0 & $-$3.9 & 2.66 & 9600 & 98 & 1184.7479 & 0.526 & 13.53 & 10.392 & 0.63 & 13.9 \\
FI Mon & M & 07 10 38.1 & $-$07 07 22 & 221.5 & 1.0 & 3.29 & 3600 & 60 & 0834.7697 & 0.930 & 12.93 & 9.632 & 0.54 & 12.2 \\
FO Cas & D & 00 17 02.6 & +60 48 10 & 118.8 & $-$1.8 & 6.80 & 14400 & 120 & 1183.5900 & 0.454 & 14.33 & 10.531 & 0.82 & 17.2 \\
GP Per & D & 04 23 19.3 & +44 14 12 & 157.9 & $-$3.8 & 2.04 & 7200 & 85 & 1186.5814 & 0.674 & 14.10 & 10.823 & 0.74 & 13.8 \\
GV Aur & D & 05 44 14.1 & +37 35 12 & 172.5 & 4.3 & 5.26 & 1800 & 42 & 0809.7890 & 0.171 & 12.09 & 8.821 & 0.58 & 12.7 \\
HQ Car & C & 10 20 32.0 & $-$61 14 58 & 285.8 & $-$3.5 & 14.07 & 3300 & 57 & 0834.8637 & 0.361 & 12.25 & 9.937 & 0.42 & 15.7 \\
HQ Per & D & 04 43 58.0 & +40 50 05 & 163.0 & $-$3.3 & 8.64 & 5400 & 73 & 0811.7616 & 0.622 & 11.61 & 8.410 & 0.59 & 13.3 \\
HW Pup & M & 07 57 42.3 & $-$27 36 07 & 244.8 & 0.8 & 13.45 & 2400 & 49 & 0831.7934 & 0.585 & 12.13 & 8.773 & 0.72 & 14.0 \\
IN Aur & D & 05 15 27.3 & +37 22 21 & 169.5 & $-$0.6 & 4.91 & 7200 & 85 & 1186.6843 & 0.266 & 13.83 & 9.854 & 0.95 & 14.8 \\
IO Cas & D & 01 47 02.8 & +59 36 23 & 129.9 & $-$2.5 & 5.60 & 10800 & 104 & 1185.5775 & 0.787 & 13.70 & 10.544 & 0.61 & 17.1 \\
NT Pup & C & 07 58 46.3 & $-$38 59 40 & 254.6 & $-$5.0 & 15.57 & 3600 & 60 & 0832.8380 & 0.792 & 12.14 & 8.134 & 0.67 & 12.0 \\
NY Cas\tablenotemark{c} & D & 00 40 23.3 & +58 37 07 & 121.5 & $-$4.2 & 4.01 & 7200 & 85 & 0809.6238 & 0.233 & 13.34 & 10.927 & 0.35 & 16.4 \\
OT Per & D & 04 38 37.9 & +47 44 24 & 157.2 & 0.6 & 26.09 & 4800 & 69 & 1185.7099 & 0.468 & 13.53 & 7.649 & 1.44 & 14.9 \\
V484 Mon & D & 06 31 05.4 & $-$02 08 48 & 212.5 & $-$5.5 & 3.14 & 14400 & 120 & 1203.6105 & 0.069 & 13.76 & 10.499 & 0.71 & 14.4 \\
WW Mon & D & 06 33 37.3 & +09 12 13 & 202.7 & 0.3 & 4.66 & 3600 & 60 & 0809.8249 & 0.399 & 12.55 & 9.510 & 0.64 & 13.6 \\
XZ CMa & C & 07 00 24.9 & $-$20 25 54 & 232.2 & $-$7.3 & 2.56 & 3600 & 60 & 0834.8182 & 0.292 & 12.76 & 10.491 & 0.27 & 13.0 \\
YZ Aur & D & 05 15 22.1 & +40 04 41 & 167.3 & 0.9 & 18.19 & 2400 & 49 & 1186.7698 & 0.219 & 10.36 & 6.717 & 0.57 & 12.1 \\
\enddata

\tablenotetext{a}{C = \citet{caldwell87}, D = DDO electronic database, M = \citet{metzger98}}
\tablenotetext{b}{S/N ratio the peak value per pixel 
in the order containing H$_\alpha$.}
\tablenotetext{c}{We corrected the observed period from 2.82d to 4.01d 
since this star is an overtone pulsator.}

\end{deluxetable}

\begin{deluxetable}{lccrclccrclccr} 
\tabletypesize{\tiny}
\rotate
\tablecolumns{14} 
\tablewidth{0pc} 
\tablecaption{Line list\label{tab:line}}
\tablehead{ 
\colhead{Wavelength(\AA)} &
\colhead{Species} &
\colhead{LEP(eV)} &
\colhead{log $gf$} &
\colhead{} &
\colhead{Wavelength(\AA)} &
\colhead{Species} &
\colhead{LEP(eV)} &
\colhead{log $gf$} &
\colhead{} &
\colhead{Wavelength(\AA)} &
\colhead{Species} &
\colhead{LEP(eV)} &
\colhead{log $gf$} 
}
\startdata
5711.09 & {\rm Mg\,{\sc i}} & 4.35 & $-$1.830 & & 5151.91 & {\rm Fe\,{\sc i}} & 1.01 & $-$3.320 & & 6355.03 & {\rm Fe\,{\sc i}} & 2.84 & $-$2.400 \\
5645.61 & {\rm Si\,{\sc i}} & 4.93 & $-$2.140 & & 5166.28 & {\rm Fe\,{\sc i}} & 0.00 & $-$4.200 & & 6393.60 & {\rm Fe\,{\sc i}} & 2.43 & $-$1.470 \\
5665.56 & {\rm Si\,{\sc i}} & 4.92 & $-$2.040 & & 5198.71 & {\rm Fe\,{\sc i}} & 2.22 & $-$2.140 & & 6408.02 & {\rm Fe\,{\sc i}} & 3.68 & $-$1.070 \\
5690.43 & {\rm Si\,{\sc i}} & 4.93 & $-$1.870 & & 5217.39 & {\rm Fe\,{\sc i}} & 3.21 & $-$1.180 & & 6411.65 & {\rm Fe\,{\sc i}} & 3.65 & $-$0.730 \\
5708.40 & {\rm Si\,{\sc i}} & 4.95 & $-$1.470 & & 5242.49 & {\rm Fe\,{\sc i}} & 3.63 & $-$0.980 & & 6430.84 & {\rm Fe\,{\sc i}} & 2.17 & $-$2.010 \\
5772.15 & {\rm Si\,{\sc i}} & 5.08 & $-$1.750 & & 5253.46 & {\rm Fe\,{\sc i}} & 3.28 & $-$1.630 & & 6494.98 & {\rm Fe\,{\sc i}} & 2.40 & $-$1.270 \\
5793.07 & {\rm Si\,{\sc i}} & 4.93 & $-$2.060 & & 5288.53 & {\rm Fe\,{\sc i}} & 3.69 & $-$1.530 & & 6518.36 & {\rm Fe\,{\sc i}} & 2.83 & $-$2.500 \\
5948.54 & {\rm Si\,{\sc i}} & 5.08 & $-$1.230 & & 5302.30 & {\rm Fe\,{\sc i}} & 3.28 & $-$0.770 & & 6574.23 & {\rm Fe\,{\sc i}} & 0.99 & $-$5.000 \\
6125.02 & {\rm Si\,{\sc i}} & 5.61 & $-$1.480 & & 5307.36 & {\rm Fe\,{\sc i}} & 1.61 & $-$2.990 & & 6575.02 & {\rm Fe\,{\sc i}} & 2.59 & $-$2.730 \\
6145.01 & {\rm Si\,{\sc i}} & 5.62 & $-$1.380 & & 5321.11 & {\rm Fe\,{\sc i}} & 4.43 & $-$1.110 & & 6581.21 & {\rm Fe\,{\sc i}} & 1.48 & $-$4.710 \\
6155.13 & {\rm Si\,{\sc i}} & 5.62 & $-$0.700 & & 5339.93 & {\rm Fe\,{\sc i}} & 3.26 & $-$0.740 & & 6592.91 & {\rm Fe\,{\sc i}} & 2.73 & $-$1.490 \\
5349.47 & {\rm Ca\,{\sc i}} & 2.71 & $-$0.310 & & 5367.48 & {\rm Fe\,{\sc i}} & 4.41 & 0.430 & & 6593.87 & {\rm Fe\,{\sc i}} & 2.43 & $-$2.420 \\
5512.98 & {\rm Ca\,{\sc i}} & 2.93 & $-$0.460 & & 5379.57 & {\rm Fe\,{\sc i}} & 3.69 & $-$1.530 & & 6609.11 & {\rm Fe\,{\sc i}} & 2.56 & $-$2.690 \\
5581.97 & {\rm Ca\,{\sc i}} & 2.52 & $-$0.560 & & 5415.19 & {\rm Fe\,{\sc i}} & 4.38 & 0.630 & & 6625.02 & {\rm Fe\,{\sc i}} & 1.01 & $-$5.370 \\
5588.76 & {\rm Ca\,{\sc i}} & 2.53 & 0.360 & & 5497.52 & {\rm Fe\,{\sc i}} & 1.01 & $-$2.850 & & 6677.99 & {\rm Fe\,{\sc i}} & 2.69 & $-$1.440 \\
5590.12 & {\rm Ca\,{\sc i}} & 2.52 & $-$0.570 & & 5506.78 & {\rm Fe\,{\sc i}} & 0.99 & $-$2.800 & & 6750.15 & {\rm Fe\,{\sc i}} & 2.42 & $-$2.620 \\
5594.47 & {\rm Ca\,{\sc i}} & 2.52 & 0.100 & & 5569.62 & {\rm Fe\,{\sc i}} & 3.41 & $-$0.540 & & 6752.70 & {\rm Fe\,{\sc i}} & 4.64 & $-$1.270 \\
5598.49 & {\rm Ca\,{\sc i}} & 2.52 & $-$0.090 & & 5586.76 & {\rm Fe\,{\sc i}} & 3.37 & $-$0.160 & & 6810.26 & {\rm Fe\,{\sc i}} & 4.60 & $-$1.000 \\
5601.28 & {\rm Ca\,{\sc i}} & 2.53 & $-$0.520 & & 5600.23 & {\rm Fe\,{\sc i}} & 4.26 & $-$1.490 & & 6945.20 & {\rm Fe\,{\sc i}} & 2.42 & $-$2.480 \\
6166.44 & {\rm Ca\,{\sc i}} & 2.52 & $-$1.140 & & 5618.63 & {\rm Fe\,{\sc i}} & 4.21 & $-$1.290 & & 7112.17 & {\rm Fe\,{\sc i}} & 2.99 & $-$3.040 \\
6449.81 & {\rm Ca\,{\sc i}} & 2.52 & $-$0.500 & & 5624.54 & {\rm Fe\,{\sc i}} & 3.41 & $-$0.800 & & 7401.69 & {\rm Fe\,{\sc i}} & 4.18 & $-$1.660 \\
6471.66 & {\rm Ca\,{\sc i}} & 2.53 & $-$0.690 & & 5701.55 & {\rm Fe\,{\sc i}} & 2.56 & $-$2.220 & & 7511.01 & {\rm Fe\,{\sc i}} & 4.18 & 0.080 \\
6493.78 & {\rm Ca\,{\sc i}} & 2.52 & $-$0.110 & & 5753.12 & {\rm Fe\,{\sc i}} & 4.26 & $-$0.710 & & 7710.36 & {\rm Fe\,{\sc i}} & 4.22 & $-$1.130 \\
6499.65 & {\rm Ca\,{\sc i}} & 2.52 & $-$0.820 & & 5775.08 & {\rm Fe\,{\sc i}} & 4.22 & $-$1.310 & & 7941.09 & {\rm Fe\,{\sc i}} & 3.27 & $-$2.330 \\
6717.69 & {\rm Ca\,{\sc i}} & 2.71 & $-$0.520 & & 5816.37 & {\rm Fe\,{\sc i}} & 4.55 & $-$0.620 & & 4993.36 & {\rm Fe\,{\sc ii}} & 2.81 & $-$3.490 \\
7148.15 & {\rm Ca\,{\sc i}} & 2.71 & 0.140 & & 5855.09 & {\rm Fe\,{\sc i}} & 4.60 & $-$1.550 & & 5100.66 & {\rm Fe\,{\sc ii}} & 2.81 & $-$4.140 \\
7202.19 & {\rm Ca\,{\sc i}} & 2.71 & $-$0.260 & & 5956.69 & {\rm Fe\,{\sc i}} & 0.86 & $-$4.610 & & 5132.67 & {\rm Fe\,{\sc ii}} & 2.81 & $-$3.900 \\
4999.50 & {\rm Ti\,{\sc i}} & 0.83 & 0.310 & & 6012.21 & {\rm Fe\,{\sc i}} & 2.22 & $-$4.070 & & 5325.55 & {\rm Fe\,{\sc ii}} & 3.22 & $-$3.220 \\
5007.21 & {\rm Ti\,{\sc i}} & 0.82 & 0.170 & & 6027.05 & {\rm Fe\,{\sc i}} & 4.07 & $-$1.110 & & 5414.07 & {\rm Fe\,{\sc ii}} & 3.22 & $-$3.750 \\
5016.16 & {\rm Ti\,{\sc i}} & 0.85 & $-$0.520 & & 6065.48 & {\rm Fe\,{\sc i}} & 2.61 & $-$1.530 & & 5425.26 & {\rm Fe\,{\sc ii}} & 3.20 & $-$3.370 \\
5024.84 & {\rm Ti\,{\sc i}} & 0.82 & $-$0.550 & & 6082.71 & {\rm Fe\,{\sc i}} & 2.22 & $-$3.570 & & 5732.72 & {\rm Fe\,{\sc ii}} & 3.38 & $-$4.670 \\
5173.74 & {\rm Ti\,{\sc i}} & 0.00 & $-$1.060 & & 6136.62 & {\rm Fe\,{\sc i}} & 2.45 & $-$1.400 & & 5991.38 & {\rm Fe\,{\sc ii}} & 3.15 & $-$3.560 \\
5210.39 & {\rm Ti\,{\sc i}} & 0.05 & $-$0.830 & & 6151.62 & {\rm Fe\,{\sc i}} & 2.17 & $-$3.300 & & 6084.11 & {\rm Fe\,{\sc ii}} & 3.20 & $-$3.810 \\
6126.22 & {\rm Ti\,{\sc i}} & 1.05 & $-$1.370 & & 6165.36 & {\rm Fe\,{\sc i}} & 4.14 & $-$1.490 & & 6149.26 & {\rm Fe\,{\sc ii}} & 3.89 & $-$2.720 \\
6258.10 & {\rm Ti\,{\sc i}} & 1.44 & $-$0.300 & & 6173.34 & {\rm Fe\,{\sc i}} & 2.22 & $-$2.880 & & 6179.38 & {\rm Fe\,{\sc ii}} & 5.57 & $-$2.600 \\
6261.11 & {\rm Ti\,{\sc i}} & 1.43 & $-$0.420 & & 6180.20 & {\rm Fe\,{\sc i}} & 2.73 & $-$2.640 & & 6247.56 & {\rm Fe\,{\sc ii}} & 3.89 & $-$2.330 \\
6554.22 & {\rm Ti\,{\sc i}} & 1.46 & $-$1.020 & & 6200.31 & {\rm Fe\,{\sc i}} & 2.61 & $-$2.440 & & 6369.46 & {\rm Fe\,{\sc ii}} & 2.89 & $-$4.250 \\
5268.62 & {\rm Ti\,{\sc ii}} & 2.60 & $-$1.620 & & 6219.28 & {\rm Fe\,{\sc i}} & 2.20 & $-$2.430 & & 6383.72 & {\rm Fe\,{\sc ii}} & 5.55 & $-$2.270 \\
5381.02 & {\rm Ti\,{\sc ii}} & 1.57 & $-$2.080 & & 6229.23 & {\rm Fe\,{\sc i}} & 2.84 & $-$2.850 & & 6416.92 & {\rm Fe\,{\sc ii}} & 3.89 & $-$2.740 \\
5418.80 & {\rm Ti\,{\sc ii}} & 1.58 & $-$1.860 & & 6230.73 & {\rm Fe\,{\sc i}} & 2.56 & $-$1.280 & & 6432.68 & {\rm Fe\,{\sc ii}} & 2.89 & $-$3.710 \\
5910.05 & {\rm Ti\,{\sc ii}} & 1.57 & $-$3.240 & & 6232.64 & {\rm Fe\,{\sc i}} & 3.65 & $-$1.280 & & 6516.08 & {\rm Fe\,{\sc ii}} & 2.89 & $-$3.450 \\
6606.95 & {\rm Ti\,{\sc ii}} & 2.06 & $-$2.790 & & 6246.32 & {\rm Fe\,{\sc i}} & 3.60 & $-$0.890 & & 7222.39 & {\rm Fe\,{\sc ii}} & 3.89 & $-$3.300 \\
7214.72 & {\rm Ti\,{\sc ii}} & 2.59 & $-$1.750 & & 6252.55 & {\rm Fe\,{\sc i}} & 2.40 & $-$1.690 & & 7479.69 & {\rm Fe\,{\sc ii}} & 3.89 & $-$3.590 \\
4924.77 & {\rm Fe\,{\sc i}} & 2.28 & $-$2.290 & & 6265.13 & {\rm Fe\,{\sc i}} & 2.17 & $-$2.550 & & 7515.83 & {\rm Fe\,{\sc ii}} & 3.90 & $-$3.430 \\
4930.31 & {\rm Fe\,{\sc i}} & 3.96 & $-$1.260 & & 6297.79 & {\rm Fe\,{\sc i}} & 2.22 & $-$2.740 & & 7711.72 & {\rm Fe\,{\sc ii}} & 3.90 & $-$2.540 \\
5014.94 & {\rm Fe\,{\sc i}} & 3.94 & $-$0.320 & & 6301.50 & {\rm Fe\,{\sc i}} & 3.65 & $-$0.770 & & 5769.06 & {\rm La\,{\sc ii}} & 1.25 & $-$0.690 \\
5044.21 & {\rm Fe\,{\sc i}} & 2.85 & $-$2.030 & & 6322.69 & {\rm Fe\,{\sc i}} & 2.59 & $-$2.430 & & 5805.77 & {\rm La\,{\sc ii}} & 0.13 & $-$1.560 \\
5049.82 & {\rm Fe\,{\sc i}} & 2.28 & $-$1.370 & & 6335.33 & {\rm Fe\,{\sc i}} & 2.20 & $-$2.190 & & 6262.29 & {\rm La\,{\sc ii}} & 0.40 & $-$1.220 \\
5083.34 & {\rm Fe\,{\sc i}} & 0.96 & $-$2.960 & & 6336.82 & {\rm Fe\,{\sc i}} & 3.68 & $-$0.920 & & 6390.48 & {\rm La\,{\sc ii}} & 0.32 & $-$1.410 \\
5141.74 & {\rm Fe\,{\sc i}} & 2.42 & $-$2.000 & & 6344.15 & {\rm Fe\,{\sc i}} & 2.43 & $-$2.920 & & 6645.13 & {\rm Eu\,{\sc ii}} & 1.38 & 0.200 \\
\enddata

\end{deluxetable}

\begin{deluxetable}{lcrccccccrc} 
\tabletypesize{\scriptsize}
\tablecolumns{11} 
\tablewidth{0pc} 
\tablecaption{Atmospheric parameters.\label{tab:param}}
\tablehead{ 
\colhead{Name} & 
\colhead{\teff (K)} & 
\colhead{$\log g$} & 
\colhead{$\xi_t$} & 
\colhead{log $\epsilon$(Fe\,{\sc i})} &
\colhead{$\sigma$} & 
\colhead{$N$} &
\colhead{log $\epsilon$(Fe\,{\sc ii})} &
\colhead{$\sigma$} & 
\colhead{$N$} &
\colhead{[Fe/H]} 
}
\startdata
CE Pup & 4975 & 0.20 & 4.05 & 7.13 & 0.15 & 26 & 7.15 & 0.04 & 4 & $-$0.40 \\
CI Per & 6000 & 1.50 & 3.35 & 6.50 & 0.26 & 23 & 6.50 & 0.24 & 8 & $-$1.04 \\
CR Ori & 5625 & 1.30 & 3.17 & 6.94 & 0.09 & 24 & 6.98 & 0.11 & 10 & $-$0.58 \\
CU Mon & 5975 & 1.30 & 2.95 & 6.97 & 0.12 & 30 & 6.99 & 0.09 & 9 & $-$0.56 \\
CY Aur & 5225 & 0.80 & 2.72 & 7.02 & 0.11 & 35 & 7.05 & 0.10 & 8 & $-$0.51 \\
EE Mon & 5900 & 1.50 & 2.77 & 7.00 & 0.10 & 33 & 7.02 & 0.17 & 9 & $-$0.53 \\
ER Aur & 5725 & 1.00 & 2.90 & 6.91 & 0.11 & 49 & 6.90 & 0.17 & 16 & $-$0.64 \\
EW Aur & 6200 & 1.30 & 2.88 & 6.61 & 0.12 & 31 & 6.68 & 0.18 & 11 & $-$0.90 \\
FI Mon & 6700 & 1.90 & 3.35 & 7.21 & 0.13 & 21 & 7.18 & 0.17 & 9 & $-$0.35 \\
FO Cas & 6300 & 1.00 & 2.84 & 6.64 & 0.14 & 40 & 6.65 & 0.17 & 17 & $-$0.90 \\
GP Per & 6400 & 1.50 & 2.69 & 6.65 & 0.15 & 38 & 6.66 & 0.15 & 15 & $-$0.89 \\
GV Aur & 6400 & 1.70 & 2.83 & 7.29 & 0.14 & 48 & 7.26 & 0.16 & 16 & $-$0.27 \\
HQ Car & 5400 & 0.80 & 3.15 & 7.11 & 0.13 & 32 & 7.16 & 0.12 & 9 & $-$0.41 \\
HQ Per & 5625 & 1.70 & 4.85 & 7.07 & 0.15 & 32 & 7.08 & 0.13 & 8 & $-$0.47 \\
HW Pup & 5750 & 1.00 & 3.09 & 7.12 & 0.11 & 26 & 7.16 & 0.14 & 10 & $-$0.40 \\
IN Aur & 6100 & 1.80 & 3.39 & 7.02 & 0.13 & 46 & 7.03 & 0.15 & 11 & $-$0.52 \\
IO Cas & 5225 & 0.30 & 3.11 & 6.75 & 0.09 & 26 & 6.73 & 0.12 & 8 & $-$0.80 \\
NT Pup & 5250 & 0.80 & 3.42 & 7.18 & 0.17 & 35 & 7.21 & 0.11 & 9 & $-$0.35 \\
NY Cas & 6050 & 1.70 & 2.62 & 6.85 & 0.14 & 45 & 6.83 & 0.18 & 18 & $-$0.70 \\
OT Per & 4700 & $-$0.30 & 5.40 & 6.64 & 0.18 & 29 & 6.69 & 0.11 & 8 & $-$0.88 \\
V484 Mon & 6550 & 1.70 & 2.98 & 7.08 & 0.11 & 23 & 7.08 & 0.15 & 11 & $-$0.46 \\
WW Mon & 6300 & 1.50 & 2.86 & 6.96 & 0.12 & 36 & 6.94 & 0.13 & 14 & $-$0.59 \\
XZ CMa & 6750 & 1.70 & 3.32 & 6.97 & 0.15 & 20 & 6.97 & 0.13 & 9 & $-$0.57 \\
YZ Aur & 5325 & 0.20 & 2.88 & 6.89 & 0.16 & 32 & 6.92 & 0.20 & 11 & $-$0.64 \\
\enddata

\end{deluxetable}

\begin{deluxetable}{lccccrccrccrccccccc} 
\tabletypesize{\tiny}
\rotate
\tablecolumns{19} 
\tablewidth{0pc} 
\tablecaption{Mean stellar abundances.\label{tab:abund}}
\tablehead{ 
\colhead{Name} & 
\colhead{[Mg/Fe]\tablenotemark{a}} & 
\colhead{[Si/Fe]} & 
\colhead{$\sigma$} & 
\colhead{$N$} & 
\colhead{[Ca/Fe]} & 
\colhead{$\sigma$} & 
\colhead{$N$} & 
\colhead{[Ti\,{\sc i}/Fe]} & 
\colhead{$\sigma$} & 
\colhead{$N$} & 
\colhead{[Ti\,{\sc ii}/Fe]} & 
\colhead{$\sigma$} & 
\colhead{$N$} & 
\colhead{[$\alpha$/Fe]} & 
\colhead{[La/Fe]} &
\colhead{$\sigma$} & 
\colhead{$N$} & 
\colhead{[Eu/Fe]\tablenotemark{a}} 
}
\startdata
CE Pup & \ldots & 0.30 & 0.17 & 10 & 0.19 & 0.06 & 2 & 0.16 & 0.16 & 6 & 0.12 & 0.00 & 2 & 0.19 & 0.43 & 0.06 & 4 & 0.45 \\
CI Per & \ldots & \ldots & \ldots & \ldots & 0.34 & 0.16 & 9 & \ldots & \ldots & \ldots & 0.25 & 0.04 & 2 & 0.30 & \ldots & \ldots & \ldots & \ldots \\
CR Ori & 0.29 & 0.28 & 0.11 & 7 & 0.22 & 0.17 & 9 & 0.25 & 0.00 & 1 & 0.10 & 0.07 & 3 & 0.23 & 0.51 & 0.05 & 4 & 0.43 \\
CU Mon & \ldots & 0.32 & 0.13 & 5 & 0.19 & 0.15 & 9 & 0.26 & \ldots & 1 & 0.19 & \ldots & 1 & 0.24 & 0.39 & 0.05 & 3 & 0.41 \\
CY Aur & 0.38 & 0.27 & 0.12 & 9 & 0.11 & 0.05 & 5 & 0.07 & 0.19 & 8 & 0.21 & 0.07 & 4 & 0.21 & 0.50 & 0.20 & 4 & 0.41 \\
EE Mon & 0.17 & 0.33 & 0.20 & 8 & 0.12 & 0.17 & 8 & 0.10 & 0.11 & 2 & $-$0.07 & 0.18 & 2 & 0.13 & 0.32 & 0.04 & 4 & 0.48 \\
ER Aur & 0.23 & 0.36 & 0.18 & 9 & 0.15 & 0.13 & 12 & 0.17 & 0.26 & 9 & 0.00 & 0.11 & 5 & 0.18 & 0.39 & 0.05 & 4 & 0.49 \\
EW Aur & 0.31 & 0.56 & \ldots & 1 & 0.27 & 0.20 & 8 & 0.33 & 0.25 & 4 & 0.28 & 0.21 & 4 & 0.35 & \ldots & \ldots & \ldots & 0.20 \\
FI Mon & 0.21 & 0.25 & 0.13 & 3 & 0.28 & 0.21 & 3 & \ldots & \ldots & \ldots & \ldots & \ldots & \ldots & 0.25 & 0.29 & \ldots & 1 & \ldots \\
FO Cas & 0.27 & 0.46 & 0.08 & 4 & 0.22 & 0.16 & 15 & 0.39 & 0.33 & 4 & 0.26 & 0.30 & 6 & 0.32 & \ldots & \ldots & \ldots & 0.40 \\
GP Per & \ldots & 0.26 & 0.06 & 2 & 0.24 & 0.20 & 9 & \ldots & \ldots & \ldots & 0.46 & 0.14 & 4 & 0.32 & \ldots & \ldots & \ldots & \ldots \\
GV Aur & 0.01 & 0.17 & 0.16 & 8 & 0.15 & 0.15 & 15 & 0.14 & 0.17 & 7 & 0.09 & 0.22 & 6 & 0.11 & 0.51 & 0.05 & 3 & 0.37 \\
HQ Car & \ldots & 0.44 & 0.18 & 9 & $-$0.04 & 0.22 & 6 & $-$0.01 & \ldots & 1 & $-$0.07 & \ldots & 1 & 0.08 & \ldots & \ldots & \ldots & 0.06 \\
HQ Per & 0.18 & 0.14 & 0.22 & 6 & 0.04 & 0.12 & 7 & 0.12 & 0.09 & 3 & 0.16 & 0.15 & 3 & 0.13 & 0.44 & \ldots & 1 & 0.52 \\
HW Pup & 0.16 & 0.44 & 0.12 & 4 & 0.09 & 0.07 & 6 & 0.06 & 0.11 & 2 & $-$0.05 & 0.01 & 2 & 0.14 & 0.39 & 0.03 & 3 & 0.27 \\
IN Aur & 0.27 & 0.24 & 0.19 & 5 & 0.12 & 0.13 & 9 & $-$0.02 & 0.05 & 5 & $-$0.08 & 0.00 & 2 & 0.11 & 0.49 & \ldots & 1 & 0.42 \\
IO Cas & 0.45 & 0.50 & 0.15 & 5 & 0.40 & 0.21 & 6 & 0.23 & 0.13 & 4 & $-$0.01 & 0.14 & 4 & 0.31 & 0.03 & 0.09 & 2 & 0.25 \\
NT Pup & 0.09 & 0.23 & 0.11 & 9 & 0.03 & 0.07 & 6 & 0.01 & 0.24 & 5 & 0.09 & 0.07 & 2 & 0.09 & 0.38 & 0.09 & 2 & 0.33 \\
NY Cas & 0.47 & 0.25 & 0.10 & 4 & 0.26 & 0.18 & 14 & 0.36 & 0.24 & 5 & 0.29 & 0.22 & 6 & 0.33 & 0.58 & \ldots & 1 & 0.55 \\
OT Per & \ldots & 0.37 & 0.14 & 8 & 0.05 & 0.21 & 5 & $-$0.01 & 0.31 & 7 & 0.11 & 0.11 & 3 & 0.13 & 0.30 & 0.15 & 3 & 0.43 \\
V484 Mon & 0.09 & 0.33 & 0.13 & 7 & 0.21 & 0.14 & 8 & 0.19 & 0.10 & 2 & 0.05 & 0.05 & 2 & 0.17 & 0.33 & \ldots & 1 & 0.51 \\
WW Mon & 0.33 & 0.33 & 0.26 & 6 & 0.23 & 0.15 & 14 & 0.10 & 0.05 & 3 & $-$0.06 & 0.05 & 4 & 0.19 & 0.23 & 0.11 & 3 & 0.32 \\
XZ CMa & \ldots & 0.37 & 0.16 & 3 & 0.14 & 0.14 & 4 & \ldots & \ldots & \ldots & 0.11 & \ldots & 1 & 0.21 & \ldots & \ldots & \ldots & \ldots \\
YZ Aur & 0.47 & 0.30 & 0.11 & 9 & 0.17 & 0.08 & 6 & $-$0.02 & 0.07 & 4 & $-$0.24 & 0.11 & 4 & 0.13 & 0.24 & 0.08 & 4 & 0.24 \\
\enddata

\tablenotetext{a}{Mg and Eu abundances were derived from 1 line.}

\end{deluxetable}

\begin{deluxetable}{lrrr} 
\tabletypesize{\normalsize}
\tablecolumns{4} 
\tablewidth{0pc} 
\tablecaption{Abundance dependences on model 
parameters for ER Aur.\label{tab:err}}
\tablehead{ 
\colhead{Species} & 
\colhead{\teff~+ 100} &
\colhead{$\log g$ + 0.3} &
\colhead{$\xi_t$ + 0.3}
}
\startdata
{\rm [Mg/Fe]} & $-$0.01 & $-$0.06 & 0.03 \\
{\rm [Si/Fe]} & 0.01 & $-$0.05 & 0.05 \\
{\rm [Ca/Fe]} & 0.03 & $-$0.05 & 0.02 \\
{\rm [Ti\,{\sc i}/Fe]} & 0.07 & $-$0.06 & 0.01 \\
{\rm [Ti\,{\sc ii}/Fe]} & $-$0.03 & 0.06 & $-$0.01 \\
{\rm [$\alpha$/Fe]} & 0.01 & $-$0.03 & 0.02 \\
{\rm [Fe\,{\sc i}/H]} & 0.13 & $-$0.02 & $-$0.05 \\
{\rm [Fe\,{\sc ii}/H]} & 0.02 & 0.10 & $-$0.06 \\
{\rm [La/Fe]} & 0.01 & 0.06 & 0.05 \\
{\rm [Eu/Fe]} & $-$0.01 & 0.06 & 0.04 \\
\enddata

\end{deluxetable}

\begin{deluxetable}{lccccrrcrrrccc} 
\tabletypesize{\tiny}
\tablecolumns{14} 
\tablewidth{0pc} 
\tablecaption{Stellar parameters and abundances for nearby
Cepheids also studied by Fry \& Carney and Andrievsky.\label{tab:fry}}
\tablehead{ 
\colhead{Name} & 
\colhead{Phase} &
\colhead{\teff} &
\colhead{$\log g$} &
\colhead{$\xi_t$} &
\colhead{[Fe/H]} &
\colhead{[Mg/Fe]} &
\colhead{[Si/Fe]} & 
\colhead{[Ca/Fe]} & 
\colhead{[Ti\,{\sc i}/Fe]} & 
\colhead{[Ti\,{\sc ii}/Fe]} & 
\colhead{[$\alpha$/Fe]} &
\colhead{[La/Fe]} &
\colhead{[Eu/Fe]} 
}
\startdata
RX Aur & 0.304 & 5400 & 1.1 & 3.30 & $-$0.27 & 0.06 & 0.24 & 0.01 & 0.11 & 0.18 & 0.12 & 0.47 & 0.42 \\
RX Aur & 0.217 & 5550 & 1.3 & 3.10 & $-$0.28 & 0.05 & 0.18 & 0.05 & \ldots & 0.11 & 0.10 & 0.50 & 0.38 \\
Del Cep & 0.399 & 5700 & 1.7 & 2.85 & $-$0.08 & $-$0.13 & 0.13 & 0.00 & 0.04 & 0.09 & 0.03 & 0.46 & 0.28 \\
Del Cep & 0.950 & 6850 & 2.2 & 2.75 & 0.05 & $-$0.05 & 0.12 & 0.06 & 0.12 & 0.04 & 0.06 & 0.36 & 0.25 \\
X Cyg & 0.432 & 4875 & 0.7 & 3.31 & $-$0.16 & 0.19 & 0.13 & 0.00 & $-$0.02 & \ldots & 0.07 & 0.34 & 0.25 \\
X Cyg & 0.615 & 4950 & 1.1 & 4.75 & $-$0.09 & \ldots & 0.02 & $-$0.15 & 0.07 & 0.09 & 0.01 & 0.39 & 0.19 \\
Zet Gem & 0.410 & 5300 & 1.5 & 3.55 & $-$0.07 & 0.12 & 0.15 & $-$0.05 & 0.05 & 0.15 & 0.08 & 0.33 & 0.27 \\
Zet Gem & 0.806 & 5700 & 1.4 & 3.05 & $-$0.07 & 0.11 & 0.18 & 0.06 & $-$0.04 & 0.02 & 0.07 & 0.39 & 0.24 \\
S Sge & 0.463 & 5550 & 1.4 & 3.16 & $-$0.11 & 0.08 & 0.18 & 0.01 & $-$0.02 & 0.07 & 0.06 & 0.25 & 0.21 \\
S Sge & 0.940 & 6400 & 1.8 & 2.77 & $-$0.08 & $-$0.03 & 0.11 & 0.02 & 0.01 & $-$0.11 & 0.00 & 0.34 & 0.23 \\
SZ Tau & 0.333 & 6000 & 1.7 & 2.51 & $-$0.04 & $-$0.03 & 0.16 & 0.05 & 0.06 & $-$0.04 & 0.04 & 0.32 & 0.19 \\
SZ Tau & 0.603 & 5725 & 1.5 & 2.45 & $-$0.15 & 0.02 & 0.12 & 0.06 & 0.01 & $-$0.05 & 0.03 & 0.28 & 0.15 \\
SV Vul & 0.283 & 5150 & 0.2 & 3.39 & $-$0.13 & 0.23 & 0.23 & 0.01 & 0.00 & 0.16 & 0.12 & 0.25 & 0.28 \\
SV Vul & 0.371 & 5050 & 0.2 & 3.28 & $-$0.11 & \ldots & 0.25 & 0.00 & $-$0.01 & 0.23 & 0.12 & 0.24 & 0.27 \\
Eta Aql & 0.955 & 6575 & 2.3 & 3.24 & 0.07 & $-$0.05 & 0.05 & 0.00 & 0.17 & 0.05 & 0.04 & 0.41 & 0.28 \\
T Mon & 0.387 & 4950 & 0.4 & 3.65 & $-$0.15 & 0.02 & 0.10 & 0.03 & $-$0.03 & 0.27 & 0.08 & 0.25 & 0.25 \\
T Vul & 0.011 & 6575 & 2.0 & 3.03 & $-$0.15 & 0.00 & 0.15 & 0.03 & 0.00 & 0.04 & 0.04 & 0.38 & 0.30 \\
U Sgr & 0.958 & 6450 & 2.1 & 3.40 & $-$0.08 & $-$0.04 & 0.00 & 0.08 & 0.06 & 0.07 & 0.03 & 0.33 & 0.18 \\
U Sgr & 0.404 & 5550 & 1.5 & 3.21 & $-$0.09 & 0.04 & 0.13 & $-$0.03 & $-$0.01 & 0.01 & 0.03 & 0.25 & 0.19 \\
\enddata


\end{deluxetable}

\clearpage

\clearpage

\clearpage

\clearpage

\clearpage

\clearpage

\clearpage

\clearpage

\clearpage

\clearpage

\clearpage

\clearpage

\clearpage

\clearpage

\clearpage

\clearpage

\clearpage

\clearpage

\clearpage

\end{document}